# Optimal Differential Astrometry for Multiconjugate Adaptive Optics. I. Astrometric Distortion Mapping using On-sky GeMS Observations of NGC 6723

Mojtaba Taheri[1] , Alan W. McConnachie[1,2] , Paolo Turri[2,3] , Davide Massari[4,5] , David Andersen[1,2], Giuseppe Bono[6,7] ,
Giuliana Fiorentino[7] , Kim Venn[1] , Jean-Pierre Véran[2], and Peter B. Stetson[2]
[1] Physics & Astronomy Department, University of Victoria, 3800 Finnerty Road, Victoria, BC V8P 5C2 Canada; mtaheri@uvic.ca
[2] NRC Herzberg Astronomy and Astrophysics, 5071 West Saanich Road, Victoria, BC V9E 2E7, Canada
[3] Department of Physics and Astronomy, University of British Columbia, 6224 Agricultural Road, Vancouver, BC V6T 1Z1, Canada
[4] INAF—Osservatorio di Astrofisica e Scienza dello Spazio di Bologna, Via Gobetti 93/3, I-40129 Bologna, Italy
[5] Kapteyn Astronomical Institute, University of Groningen, NL-9747 AD Groningen, Netherlands
[6] Dipartimento di Fisica, Università di Roma Tor Vergata, via della Ricerca Scientifica 1, I-00133 Roma, Italy
[7] INAF—Osservatorio Astronomico di Roma, via Frascati 33, I-0040 Monte Porzio Catone, Italy
Received 2020 August 10; revised 2022 February 13; accepted 2022 February 15; published 2022 March 23

## Abstract

The Extremely Large Telescope and the Thirty Meter Telescope will use state of the art multiconjugate adaptive optics (MCAO) systems to obtain the full $D^4$ advantage that their apertures can provide. However, to reach the full astrometric potential of these facilities for on-sky science requires understanding any residual astrometric distortions from these systems and find ways to measure and eliminate them. In this work, we use Gemini multiconjugate adaptive optic system (GeMS) observations of the core of NGC 6723 to better understand the on-sky astrometric performance of MCAO. We develop new methods to measure the astrometric distortion fields of the observing system, which probe the distortion at the highest possible spatial resolution. We also describe methods for examining the time-variable and static components of the astrometric distortion. When applied to the GeMS Gemini South Adaptive Optics Imager (GSAOI) data, we are able to see the effect of the field rotator at the subpixel level, and we are able to empirically derive the distortion due to the optical design of GeMS/GSAOI. We argue that the resulting distortion maps are a valuable tool to measure and monitor the on-sky astrometric performance of future instrumentation. Our overall astrometry pipeline produces high-quality proper motions with an uncertainty floor of ∼45 μas yr⁻¹. We measure the proper motion dispersion profile of NGC 6723 from a radius of ∼10″ out to ∼1′ based on ∼12,000 stars. We also produce a high-quality optical–near infrared color–magnitude diagram, which clearly shows the extreme horizontal branch and main-sequence knee of this cluster.

*Unified Astronomy Thesaurus concepts:* Globular star clusters (656); Astrometry (80); Astronomical instrumentation (799); Proper motions (1295); Ground telescopes (687); Astronomical seeing (92); Astronomical optics (88)

*Supporting material:* animation

## 1. Introduction

A new generation of extremely large ground-based telescopes—the Thirty Meter Telescope (TMT; Sanders 2013), the Extremely Large Telescope (ELT; Ramsay et al. 2014), and the Giant Magellan Telescope (Johns et al. 2012)—are in active development. These state-of-the art facilities will revolutionize astronomy thanks to their very large apertures, which simultaneously increase their light gathering ability and their resolving power. For their resolving power, the key technology that enables these advances is adaptive optics (AO), which compensates for the disturbing effects that the atmosphere has on the resolving power of ground-based telescopes.

AO systems come in different configurations with different levels of complexity, depending on the application. A key configuration that is critical for both the TMT and ELT is multiconjugate adaptive optics (MCAO). The desirable aspect of this architecture is its ability to provide uniform correction across a (relatively wide) one arcminute field of view, to be compared to a typical classic AO system that provides efficient correction across a field of view of the order of a few arcseconds (Neichel & Lu 2014). In a single-conjugate AO (SCAO) system, a single wave front measurement is made, and a single deformable mirror (DM) corrects for the cumulative effect of all the layers of turbulence in the cylinder of atmosphere between the telescope and the science target. The size of the corrected field is typically ∼1′ and is characterized by the "isoplanatic angle," beyond which the rest of the telescope field of view is uncorrected. MCAO systems provide a much wider corrected field of view by using multiple wave front sensors to provide a 3D tomographic-wave front measurement of the cone-shaped volume of atmosphere in front of the telescope. The wave front is corrected by using multiple DMs (at least two), which are conjugated to turbulence layers at different altitudes, and this provides relatively uniform correction across a wide field of view.

The only MCAO instrument that is currently in regular science operation is at the Gemini South telescope. The Gemini multiconjugate adaptive optic system (GeMS) and the Gemini South Adaptive Optic Imager (GSAOI) are instruments placed on the Cassegrain focus of this 8 m facility (Carrasco et al. 2012). GeMS uses a constellation of five laser guide stars (LGSs) combined with three natural guide stars (NGSs). The LGS constellation provides tomographic information about the turbulence layer in the cone-like volume of atmosphere that is in front of the primary mirror of the telescope. However, LGSs are not sufficient to remove all phase aberration modes, as they are blind to low-order phase aberration modes. This causes the "tilt anisoplanatism" problem, which is well described and addressed in Flicker et al. (2016). Here, NGSs come to help for measuring and calibrating low-order phase aberration modes like tilt/tip and focus. The real-time controller then processes the information gathered from the 5LGS+3NGS to create a tomographic map of the turbulence. This tomographic map provides the necessary information to drive two deformable mirrors conjugated altitudes of 0/9 km in the atmosphere. These mirrors provide tomographic-wave front adjustment, which provides the wide-field correction advantage in addition to a uniform wave front correction across the field of view (d'Orgeville et al. 2012; Rigaut et al. 2012).

The relatively wide, corrected, fields of view of the next generation of MCAO telescopes makes them ideal facilities to study the dynamics of dense stellar fields like the core of globular clusters (Saracino et al. 2016; Fiorentino et al. 2016; Dalessandro et al. 2016; Miller et al. 2019). Critical to this science is the determination of residual astrometric errors that act to reduce the precision of astrometric data gathered by these systems, in order to develop methods to correct for them. Our group has previously examined GeMS data to determine the optimal methods for obtaining precision stellar photometry that takes into account the variation in the structure of the point-spread function resulting from the MCAO performance (Turri et al. 2015, 2017; Monty et al. 2020). Massari et al. (2016b) also looked into the astrometric limits of GeMS/GSAOI in tandem with the Hubble Space Telescope (HST) and compared the results to the HST–HST proper motion measurements. Here, we use similar data sets to understand the global distortions of the MCAO system and their impact on science-based astrometric measurements across the field.

In astrometric studies, the principal scientific measurement is that of the position of stars (or other astronomical objects) in each observation/epoch relative to some invariable frame of reference. Although the array detectors used for these observations have linear spatial geometry, other factors such as optical design imperfections, telescope structure flexure under a varying gravity vector, and the AO performance (Neichel & Lu 2014; Riechert et al. 2018; Patti & Fiorentino 2019) can distort the linearity of the ideal field of view and degrade the precision of astrometric measurements. The resulting distortion field can typically be described by a complicated vector field that could be significantly time-variable depending on the instrument structure design. For example, in the case of Gemini South and the GeMS/GSAOI system, the time-variable changes are more significant due to the Cassegrain mounting of the AO system and the imager. The Cassegrain mounting increases the structure flexure effect caused by the varying gravity vector and eventually causes time-variable field distortion. This will be less of a problem for TMT and ELT as their designs mount these systems on the significantly more stable Nasmyth platform (Ramsay et al. 2014; Larkin et al. 2016).

In addition to flexure, another important factor that can contribute to the intensity of the distortion field is the optical design of the instrument. For example, the number of off-axis parabola (OAP) mirror relay pairs are an important design factor that affects the distortion field, as it is necessary to use two pair of OAPs in order to simultaneously eliminate phase and field distortion. However, GeMS only uses one pair of OAPs (Patti & Fiorentino 2019). This causes the instrument to suffer from a field distortion component induced by the optical design. This is an important design lesson from GeMS that is now considered in the optical design of the TMT and ELT (Atwood et al. 2010). The net intensity of the overall distortion field caused by different factors can be much larger than the actual proper motion that is being measured for GeMS/GSAOI, as we will demonstrate later. Therefore, it is essential to develop methods to efficiently compensate for the distortion field of current and future generations of MCAO systems, in order to perform astrometry-focused science. We note that part of this methodology may include developing techniques that allow for the direct, on-sky, measurement of any remaining distortions in the optical system simultaneous to the science observations.

There have been many prior efforts to characterize and correct for the field distortion of AO instruments and improve the resulting astrometry. Some of these efforts are based on decomposing the field distortion into different components. Cameron et al. (2009) was one of the first studies that efficiently decomposed different components of field distortion and astronomical uncertainties for an AO-assisted observation. This paper provided a complete analysis of the astrometric precision for a data set that was taken using the Hale 200 inch telescope equipped with a classical AO system. In this study, they found that differential tilt jitter is one of the most significant sources that degrades the precision of astrometric measurement. They also showed that it is possible to reach an astrometric precision of the order of 100 $\mu$as using three 2 minute observations for narrow angular separation and bright stars. A similar study was performed by Fritz et al. (2010) for the crowded field of the Galactic Center, which reached similar conclusions.

However, these results are not necessarily valid for other flavors of AO systems. In MCAO devices, the tilt jitter effect is different compared to single-conjugate systems as multiple guide stars are being used. The field of view of the observation is also significantly wider (arcminutes rather than arcseconds). Fritz et al. (2015) conducted one of the first studies for GeMS and examined the absolute astrometric accuracy using background galaxies as a reference frame. They concluded that GeMS has an overall astrometric accuracy of the order of 400 $\mu$as. Fritz et al. (2015) also provide a relatively detailed astrometric error budget for GeMS observations, and similar efforts for characterizing the error budget for an MCAO system have been done by Schöck et al. (2014). These authors evaluated a comprehensive error budget for the Narrow-field Infrared Adaptive Optics System Infrared Imaging Spectrograph (the analogous instruments to GeMS/GSAOI for TMT; see Larkin et al. 2016; Crane et al. 2018). Based on this study, the expected differential astrometric precision for TMT in crowded narrow fields in the $K$ band could be as small as 37 $\mu$as. Of course, results of this study have yet to receive on-sky validation.

Another approach for measuring the distortion field and improving the astrometric performance of AO systems is to use an internal calibration grid. Here, internal grid pinholes are illuminated to provide a reference measurement to study the distortion of the AO system. In this approach, the net distortion caused by the internal instrument optics is measured and can be compensated for. This method was originally considered for use in GeMS/GSAOI, and results of initial tests of the calibration grid are reported in Riechert et al. (2018). However, we understand that it has not seen regular use in science operations thus far. Riechert et al. (2018) suggests that distortion calibration can be done across the field of view with a precision of 170 μas rms. More sophisticated technologies to fabricate and implement calibration masks are being considered for future improvements of GeMS (Dennison et al. 2016) and large aperture telescopes like TMT (Service et al. 2019). Such techniques are more efficient in comparison to on-sky measurements considering the demands on telescope time, but by design they do not consider the entire system simultaneously. Cross-check and validation of these measurements with on-sky results will remain an essential component of astrometric studies.

In the absence of a precise distortion model for the relevant ground-based MCAO instrument, the main approach used to perform high-precision astrometry is to calibrate to on-sky observations that are "distortion-free". Typically, HST observations are used as this instrument and telescope are extremely well calibrated and the field distortions are well understood, and this approach has been used heavily with GeMS observations (Anderson & King 2003; Bellini et al. 2011b; Massari et al. 2016b; Dalessandro et al. 2016; Ammons et al. 2016; Fritz et al. 2017). Of course, in planning for the future, it is necessary to move toward precise methods that do not rely heavily on auxiliary observations (especially expensive space observations), as it is just not practical to be dependent on space-based imaging for ground-based AO observations. It is with an eye to this era that we undertake the present study.

The work described in this paper is one of the first steps to move toward independent, high-precision astrometry for the new generation of giant telescopes. We present our methodology for the analysis, measurement and removal of astrometric distortions for GeMS, using HST Advanced Camera for Surveys (ACS) as an auxiliary, first-epoch data set. This is a necessary, intermediate step, moving toward our ultimate goal of enabling precision proper motions from dual epoch MCAO ground-based observations. As part of this analysis, we show how we can use our GeMS/GSAOI observations to make empirical measurements of the optical distortion of the system, which could be of use in understanding the behavior of the system over time.

Our target observations are of the globular cluster NGC 6723, taken as part of the observational program first presented in Massari et al. (2016a) and Turri et al. (2015; GS-2013A-Q-16, PI: A. McConnachie). All globular clusters used as part of this program also have HST/ACS observations (Sarajedini et al. 2007). The short-term goal of this program is to study the space motion and stellar content of Galactic satellites seen with GeMS/GSAOI, and the long-term goal is to prepare for the exciting photometric and astrometric potential of the extremely large telescopes.

The observational details, the primary data reduction, and the photometric analysis are described in Section 2. In Section 3, we described the details of the astrometric analysis. Section 4 discusses the resulting distortion maps and relates these to their origins in the observing system (telescope plus instruments). Section 5 presents our processed astrometric and photometric data set for the globular cluster, including the tangential velocity dispersion profile and the color–magnitude diagram (CMD), and compares our results to the literature. Section 6 summarizes and concludes. The next paper in this series will present a stellar populations and dynamical analysis of NGC 6723 based on the CMD and velocity dispersion profile derived here.

## 2. Data Acquisition and Preparation

### 2.1. Observations

Our analysis is based on data from the Gemini South GeMS/GSAOI, the only MCAO imager system in regular science operation. GSAOI consists of $2 \times 2$ Rockwell HAWAII-2RG arrays arranged in a square configuration. Each chip has $2048 \times 2048$ pixels with a total field of $85'' \times 85''$ and a pixel scale of $0.''02$ pixel$^{-1}$. Chips are separated by a 2 mm gap. Figure 1 shows the geometry of the GSAOI chips, including their numbering scheme.

Our observations took place on 2013 April 18 7:55 to 9:28 UTC and consisted of $8 \times 160$, $1 \times 90$, and $1 \times 25.5$ s exposures in each of the $K_S$ and $J$ bands. Table 1 shows the observational log. For this study, we only use the $K_S$ band observation, which benefits from the better AO correction and provides better astrometric precision compared to the $J$ band. The average airmass for the $K_S$ 160 s observations is 1.05. Exposures are dithered in a grid pattern, which fills in the gaps between detector chips and also improves the photometric precision. The grid pattern offset step is ∼3″ with a maximum offset of 5.′6 in each axis. Dither positions are shown in Figure 1, which also shows the positions of the LGSs and NGSs.

In Figure 1, the background image is an HST/ACS image that was observed as part of the ACS Survey of Galactic Globular Clusters (Sarajedini et al. 2007). We use this catalog as the first-epoch measurements, which provides us with a temporal baseline of 6.75 yr. The photometric and astrometric reduction process of this catalog is explained in detail in Anderson et al. (2008). Specifically, these authors identify stars simultaneously in the multiple dithered exposures for each cluster and measure their positions and magnitudes using the best available point-spread function models, correcting for distortion and placing the stars on an astrometric reference frame that is crossmatched to the Two Micron All Sky Survey (2MASS). The publicly available catalog of HST/ACS data that we use[8] contains two parameters, xsig and ysig, that correspond to the expected astrometric measurement error in units of 50 mas pixels. These were derived based on the average position of each star in the ∼4 F606W and ∼4 F814W observations. An rms about that average was determined, such that xsig, ysig correspond to rms/$\sqrt{8}$ (J. Anderson, *private communication*). We adopt these as the random uncertainties on the positions of the stars in the HST catalog. As discussed in Anderson et al. (2008), the remaining distortion-related positional uncertainties are at the level of ∼0.01 pixels (0.5 mas) across the field.

### 2.2. Data Preparation

We follow the method described in Turri et al. (2017) for data preparation and photometric reduction. This method is carefully tailored for the GeMS/GSAOI observations. In brief, we use the IRAF Gemini package for performing primary data reduction. For flat-fielding, we combine the dome and twilight flats for each chip individually. The dome flat provides high signal-to-noise information; however, its spectral response is different than that of the twilight flat. We combine the two by applying a high-pass filter on the dome flat to exclude high-spatial resolution features that may be caused by the dome flat acquisition mechanism. A $3\sigma$ clipping and a median are then used to add the information from the twilight flat. The output of this process is a "super flat" frame that is used for flat-fielding raw exposures.

---

[8] https://archive.stsci.edu/prepds/acsggct/

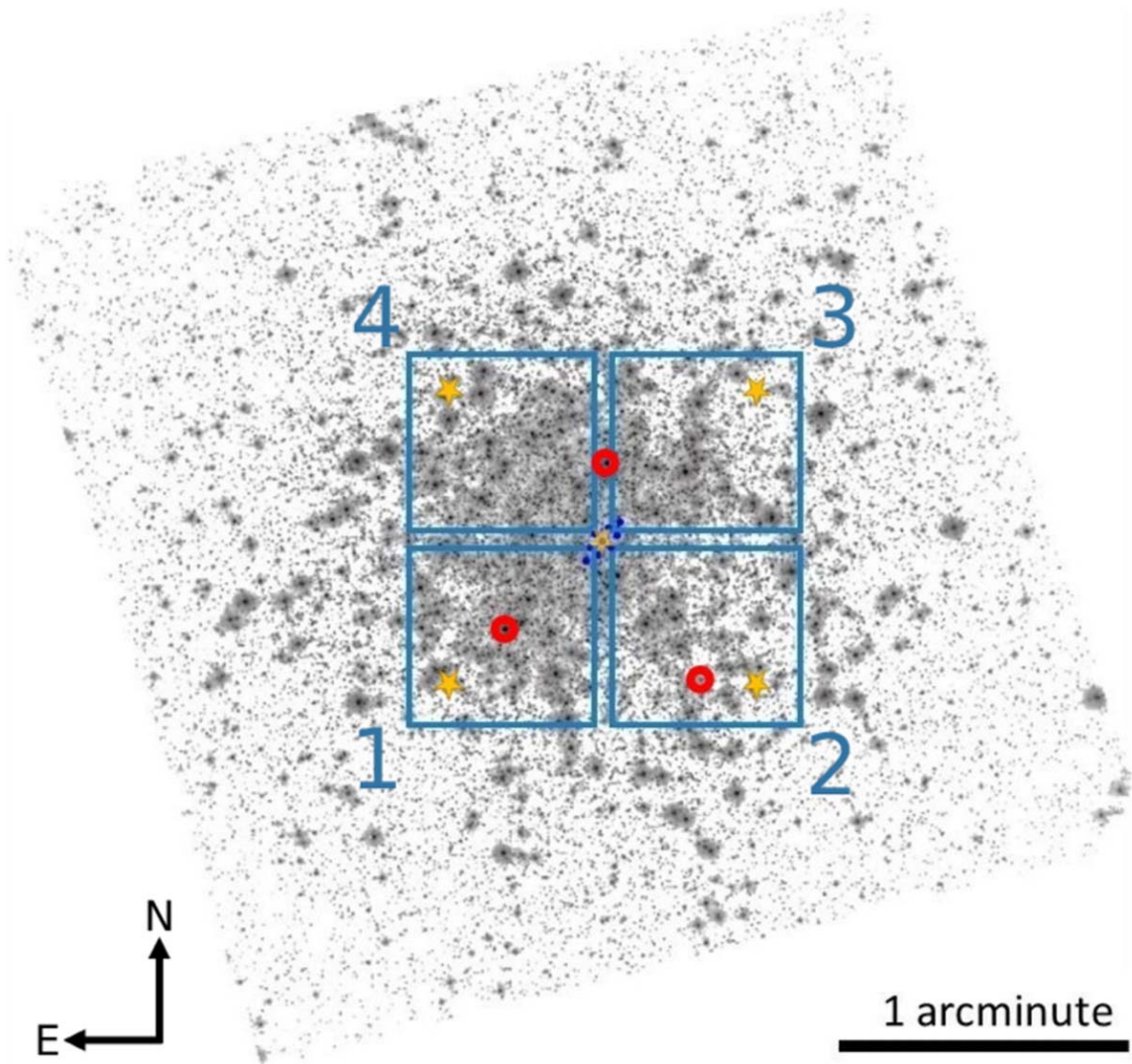

**Figure 1.** HST/ACS image from Sarajedini et al. (2007). Overlaid as blue squares is the field of view of GeMS/GSAOI. The dark blue constellation of points shows the set of field centers used for our set of dithered subexposures. The red circles indicate the positions of the NGSs, and the five yellow stars represent the positions of the LGSs.

The GSAOI detector also suffers from some dead and bright pixels. The Gemini IRAF package provides a map of these pixels and removes them from the image, preventing their destructive effect on the accuracy of the photometry. We correct for the nonlinearity of each detector chip by applying a quadratic polynomial correction measured during the GSAOI commissioning process (Carrasco et al. 2012). We do not use dark frames as the dark current is ~0.01 e s$^{-1}$ px$^{-1}$, which is totally negligible for the exposure times we used. We also did not use sky frames, as this only introduces an additional systematic uncertainty into the photometry. Instead, we evaluate the sky background at the position of each star by measuring the sigma-clipped median value of pixels in an annulus surrounding each of them. Figure 2 shows a before-and-after image of part of the GSAOI field following all these procedures.

### 2.3. Photometric Reduction

High-accuracy astrometry on multiepoch data requires determining the precise positions of each star. To achieve this, we use a careful PSF photometric analysis using the DAOPHOT II (Stetson 1987) suite of programs. The use of two discrete DMs conjugate to two different altitudes to correct the full 3D volume of turbulence above the Gemini telescope will lead to field-dependent wave front errors that cause spatial variations in the PSF (Neichel & Lu 2014; Turri et al. 2015; Massari et al. 2016b). This must be carefully considered while performing any kind of photometric analysis and was the focus of earlier contributions in Turri et al. (2017). Critically, DAOPHOT II can vary the PSF model for stars across the field using either a bilinear or cubic model, which makes it an ideal tool for analyzing MCAO and GeMS data.

The first step of the photometric analysis is to find the stars using Gaussian profile convolution. All star-like objects in the field are filtered by their sharpness and roundness parameters. This helps to distinguish stars from cosmic rays, extended objects and other artifacts. The next step, based on Turri et al. (2017), is to combine all exposures to create a master frame. This practice is primarily to reach fainter magnitudes, critical for accurate and deep photometry. However, we do not recommended this for astrometric analysis where working with the raw positions per exposure reduces systematic errors, at least prior to correction of the astrometric distortion. For the current analysis, we therefore combine information from all frames after individual distortion corrections on each exposure. Thus, in contrast with Turri et al. (2017), the rest of the photometric analysis summarized below is performed on each individual exposure and each chip.

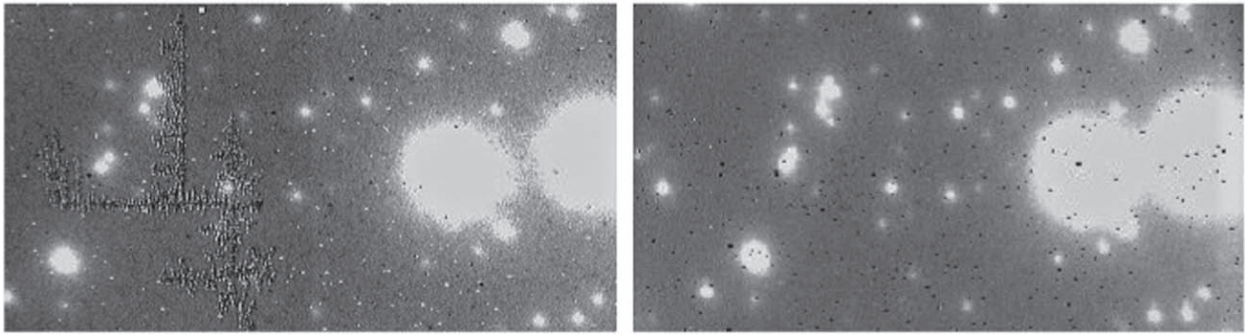

**Figure 2.** Small region of chip number 1 before (left) and after (right) primary data processing. See text for details.

**Table 1**
Observing Log for the GeMS/GSAOI Observation

| UT Time | Band | Exposure Time (sec) | Airmass | Exposure Reference |
|---|---|---|---|---|
| 2013-04-18 8:50 | $J$ | 21.5 | 1.03 | |
| 2013-04-18 8:51 | $J$ | 90 | 1.03 | |
| 2013-04-18 8:56 | $J$ | 160 | 1.03 | |
| 2013-04-18 9:00 | $J$ | 160 | 1.03 | |
| 2013-04-18 9:03 | $J$ | 160 | 1.02 | |
| 2013-04-18 9:06 | $J$ | 160 | 1.02 | |
| 2013-04-18 9:10 | $J$ | 160 | 1.02 | |
| 2013-04-18 9:13 | $J$ | 160 | 1.02 | |
| 2013-04-18 9:17 | $J$ | 160 | 1.02 | |
| 2013-04-18 9:20 | $J$ | 160 | 1.01 | |
| 2013-04-18 8:00 | $K_\mathrm{S}$ | 21.5 | 1.1 | |
| 2013-04-18 8:01 | $K_\mathrm{S}$ | 90 | 1.1 | |
| 2013-04-18 8:05 | $K_\mathrm{S}$ | 160 | 1.09 | #1 |
| 2013-04-18 8:17 | $K_\mathrm{S}$ | 160 | 1.07 | #2 |
| 2013-04-18 8:27 | $K_\mathrm{S}$ | 160 | 1.06 | #3 |
| 2013-04-18 8:31 | $K_\mathrm{S}$ | 160 | 1.05 | #4 |
| 2013-04-18 8:34 | $K_\mathrm{S}$ | 160 | 1.05 | #5 |
| 2013-04-18 8:38 | $K_\mathrm{S}$ | 160 | 1.05 | #6 |
| 2013-04-18 8:41 | $K_\mathrm{S}$ | 160 | 1.04 | #7 |
| 2013-04-18 8:45 | $K_\mathrm{S}$ | 160 | 1.04 | #8 |

The next step in the photometric analysis is choosing suitable stars to provide a PSF model and calculating the lookup table that determines the variation of the PSF across the field. Choosing stars for PSF fitting is typically accomplished by using the PICK command in the DAOPHOT II suite. The PICK command tries to delineate the number of stars across the field that are bright and isolated enough to be used for PSF modeling. We noticed that, in some very crowded regions of the field, the PICK command fails to find enough sufficiently isolated stars. Selecting a large number of isolated stars in a dense field is challenging but necessary. It ensures that the effect of photon noise and other artifacts like bad pixels on the final PSF profile are minimized and the lookup table has enough data points to efficiently interpolate the whole field of view. Therefore, we developed a script specifically designed to pick stars from more crowded fields. This script uses Gaussian fitting and the average distance between stars to ensure that the picked star profile is not significantly disturbed by neighboring bright objects. Eventually, we chose 100 stars for each chip. This process is repeated for each exposure and each chip separately.

We used the Lorentzian function for the analytic part of the PSF model. It is typically the best match for the AO PSF (Drummond 2012). We also choose cubic interpolation for modeling the lookup table that corrects for the variation of the PSF across the field. The eventual output of the PSF photometry procedure is a catalog containing the precise position and instrumental magnitude of each star for each exposure and each chip. This catalog is used in the rest of this paper to represent star positions measured by GeMS/GSAOI.

### 2.4. Photometric Calibration

The last step in the photometric analysis is the photometric calibration. We calibrate the photometric data using the 2MASS catalog (Cutri et al. 2003; Skrutskie et al. 2006).

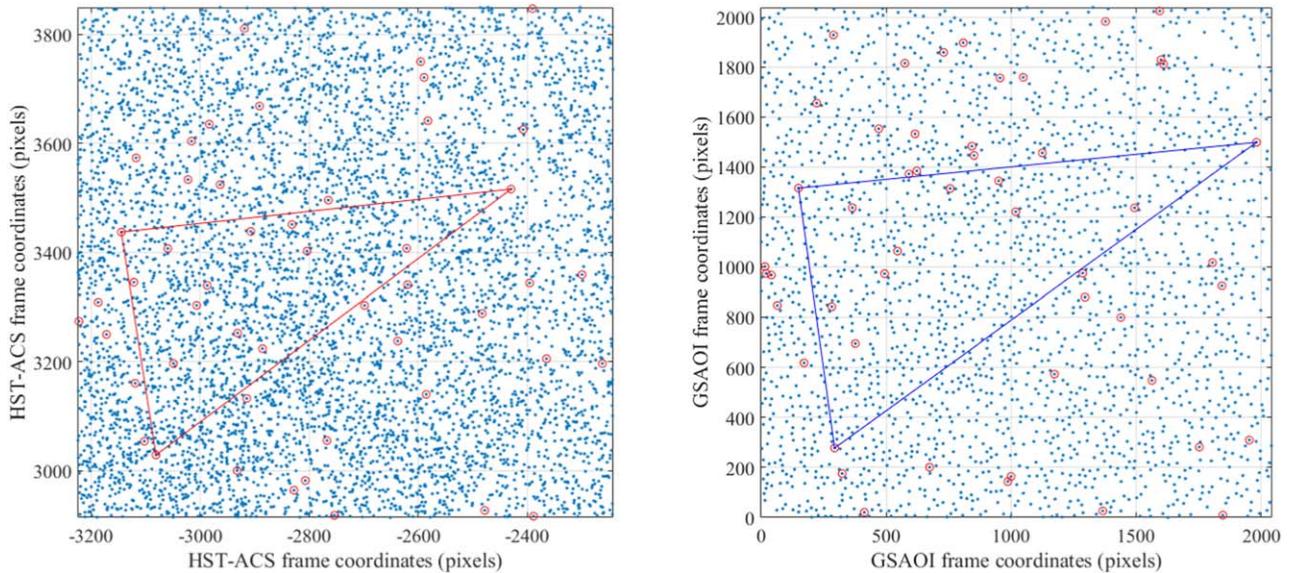

**Figure 3.** Each blue point represents a star in the HST catalog (left panel) and GeMS chip number 4 (right panel). These subcatalogs have already been filtered to ensure they do not have any very nearby neighbors, to avoid unnecessary confusion. The brightest stars of each subcatalog are indicated by red circles. One of the matched triangles between the two data sets is plotted: each matched triangle provides three votes for the three pairs of stars. The origin of the coordination system and the scales in $x$ and $y$ between these two panels are totally different, but it is clear that by finding many of these matched triangles it is possible to robustly find matching stars between the two catalogs.

The process of matching stars between our catalog and the 2MASS photometric reference is challenging. First, the resolution of the two catalogs is very different, and this causes multiple resolved stars in our catalog to be seen as one star in the 2MASS catalog. This is discussed at length in Turri et al. (2017), and they correct explicitly for this effect. We do not require the same photometric precision as Turri et al. (2017), and we find good results from hand-picking the most isolated stars in both catalogs. Second, the depth of the two catalogs is quite different, which makes finding common nonsaturated stars very difficult. We checked each pair of stars by eye for the final calibration process. After verifying the matched stars, we use a maximum-likelihood estimator (MLE) algorithm to find the best estimates of the zero points and the uncertainty of the calibration for each chip. Table 2 shows the results of this process.

## 3. Astrometric Analysis

### 3.1. Overview

While AO technology helps compensate for the degrading effect of the Earth's atmosphere on the resolving power of telescopes, MCAO observations can suffer from field distortion. Field distortion can be described as a vector field of the difference between the real (relative) positions of astronomical objects and what the instrument records for these objects across the field of view. Precise knowledge of the field distortion for a specific observation/instrument allows astrometric measurements to be made with greatly reduced systematic errors.

It is important to compensate for the relative distortion fields between our two epochs, specifically the AO-based catalog and the HST/ACS catalog. However, the fact that these are taken at two different epochs is a complicating factor. While this allows for the measurement of the proper motion of stars, it creates a new challenge in our ability to distinguish between the displacements in position due to the telescope/instrument distortion and displacements due to actual physical movement of the stars. An additional factor that is important to consider is that not all sources present in one catalog are present in the other, and indeed our initial AO-based catalogs include all the usual spurious detections that are obtained in single-frame photometric analysis (e.g., additional sources in the halos of bright stars). However, the methodology that we develop is robust to these effects and indeed will be shown to be highly effective at removing spurious sources.

**Table 2**
$K_s$−band Zero-point Magnitudes for each GSAOI Chip, Calibrated to the 2MASS Catalog

| Chip# | Zero-point (GSAOI-2MASS) | Zero-point Uncertainty |
|---|---|---|
| 1 | −5.33 | 0.02 |
| 2 | −5.62 | 0.04 |
| 3 | −5.20 | 0.02 |
| 4 | −5.01 | 0.03 |

The astrometric pipeline we have developed exploits specific characteristics of instrumental distortion maps to separate these two components. The main distinction between the distortion component and the proper motion component is the difference in the statistical spatial frequency of the displacement vector field (DVF). The DVF is the vector field created by spatially connecting the same stars between the two different observations that have been matched to each other using only the shift, scale, and rotation. The distortion map intrinsically has a continuous nature with low-spatial frequencies changing across the field of view for each detector chip. This is in contrast with the relative proper motion component, which is statistically dominated by random higher-spatial frequency behavior. By modeling the distortion field based on the low-spatial-frequency behavior of the DVF, we capture the majority of the distortion field component without losing any significant proper motion information. Our pipeline measures and models the low-spatial-frequency behavior of the DVF by analyzing local neighborhoods of vectors and eventually builds the field distortion model based on this information. We now present this method in more detail.

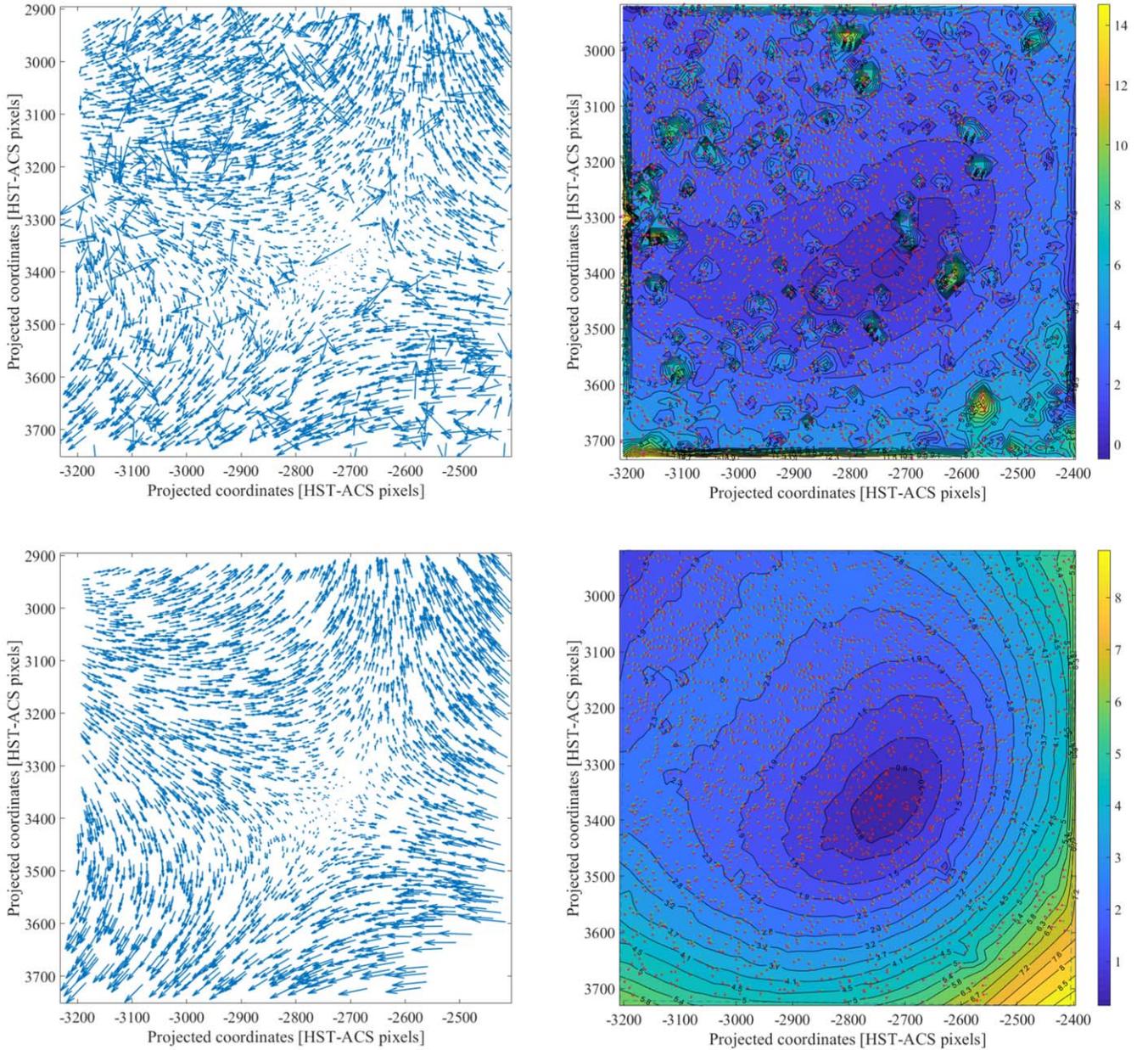

**Figure 4.** Top: initial DVF for exposure 1 and chip 4, prior to the first loop cycle of the zonal analysis. The left panel shows the vector field, and the right panel shows the magnitude. Each red–green pairs of dots in the right panel represent one pair of stars matched between the two catalogs. There are 3251 star pairs with an rms displacement of 167 mas in this map. The spatial unit of the color bar is HST/ACS pixels (∼50 mas). Bottom: same as top panel after the first loop cycle. 2502 star pairs with an rms displacement of 135 mas remained in the field. Note that only low-spatial-frequency components remain, and all higher frequency components have been removed.

### 3.2. Finding Common Stars between Catalogs

After performing the photometric analysis described in the previous section, we produce a catalog of precise positions of detected objects for each chip of each exposure. Our first step is to do a preliminary match of each of these catalogs to the HST/ACS catalog that gives us our first-epoch measurements.

We have written a vote-based star pattern matching algorithm to do this match, which works for two catalogs that have completely different coordinate systems, in terms of offset, scale, and rotation. We initially use a small subset of each catalog consisting of the highest-luminosity members from each catalog, in order to have a good chance of having common stars in each catalog and to optimize the processing speed of the algorithm. We also filter both subsets to remove stars with very nearby neighbors in order to reduce the probability of incorrect matches.

The two subcatalogs feed into a voting-based star pattern matching module. This module uses angles of triangles that can be formed in each subcatalog as a feature to find match candidates between the two catalogs. For any three stars in either of the subcatalogs, we form a triangle (see Figure 3) and compare it to the triangles formed in the other subcatalog. If the difference between the angles of the triangles in both subcatalogs is below a certain threshold, then this is recorded as a matched triangle pair. This creates three "votes" for the three pairs of stars contributing to the triangles. Many triangles are considered and compared, and votes accumulate in a votes matrix, where element $n_{i,j}$ represents the number of votes for the $i$th star in one subcatalog being matched to the $j$th star in the other subcatalog. The end result is that the votes matrix reveals the best candidates in the two subcatalogs to be the same stars. Figure 3 shows an example of matched triangles between the two data sets.

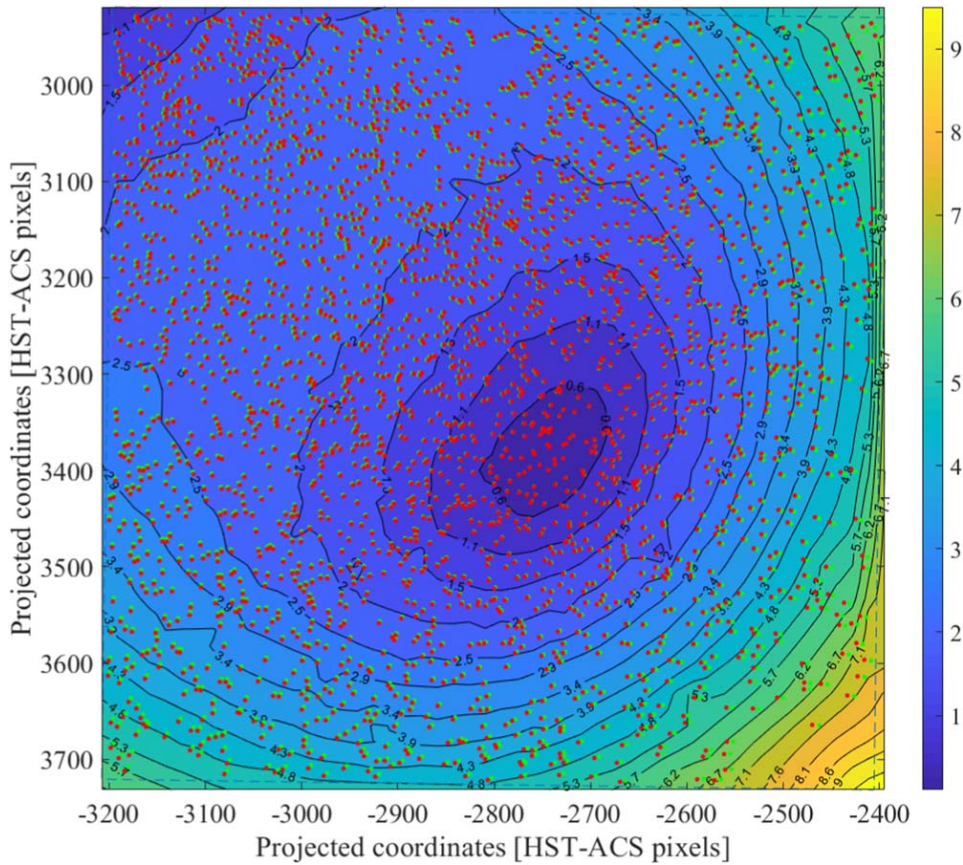

**Figure 5.** The final displacement map after the completion of the second loop cycle, for the same exposure as Figure 4. After the second loop cycle, 79 more pairs of stars are recovered, and the rms is reduced by 5 mas compared to the end of the first loop cycle. Each GSAOI pixel is equivalent to 0″.02.

### 3.3. Coordinate Transformation

The previous step has identified common stars between the two catalogs, which we can now use to define the appropriate transformations between catalogs. While the vote-based star pattern matching method is very robust against false-positive matches, we add a second stage analysis to increase the robustness even more. In particular, we calculate all possible transformations (correcting for offset, scale, and rotation) for every combination of four matched pairs. The star pairs that we use are the highest-voted matches between the two catalogs found in the voting process. Assuming the two catalogs are not reflected relative to each other, each similarity transformation is described by four parameters; two degrees of freedom for $x$ and $y$ translation, scale, and rotation. Any false-positive match that passes through this process from the previous step will cause a family of wrong and inconsistent transformations. However, all combinations of correct four pair sets will produce a unique, correct transformation. Repeated *correct* transformations will result in the same set of four parameters "piling up", whereas any incorrect transformations will scatter throughout the 4D space of all possible transformations. We use a 4D histogram analysis to recognize the most-repeated transformations between all the possible ones. This method is particularly resistant against wrongly identified pairs that may pass through the previous step. We then only use those star pairs that contributed to these correct transformations and calculate the final similarity transformation between the two catalogs based on these stars (by minimizing the rms scatter). Using this transformation, we convert all the positions in both the GSAOI catalog and the HST catalog into the same frame of reference. Stars in the transformed GSAOI positions are matched to their HST counterparts based primarily on their spatial proximity in this new frame of reference. Using all matched pairs, we are able to produce the first version of the DVF.

### 3.4. Calculating the Distortion Model

The initial DVF for chip 4 of our observation using the match between the GSAOI and HST catalogs described in the previous subsection is presented in the top row of Figure 4. This DVF contains all raw displacements measured between the two epochs, including distortion effects, proper motions, and incorrect matches (either between stars that do not actually match or including "stars" that are actually spurious detections). We remind the reader that these diagrams have already been corrected for shift, scale, and rotation using the similarity transformation.

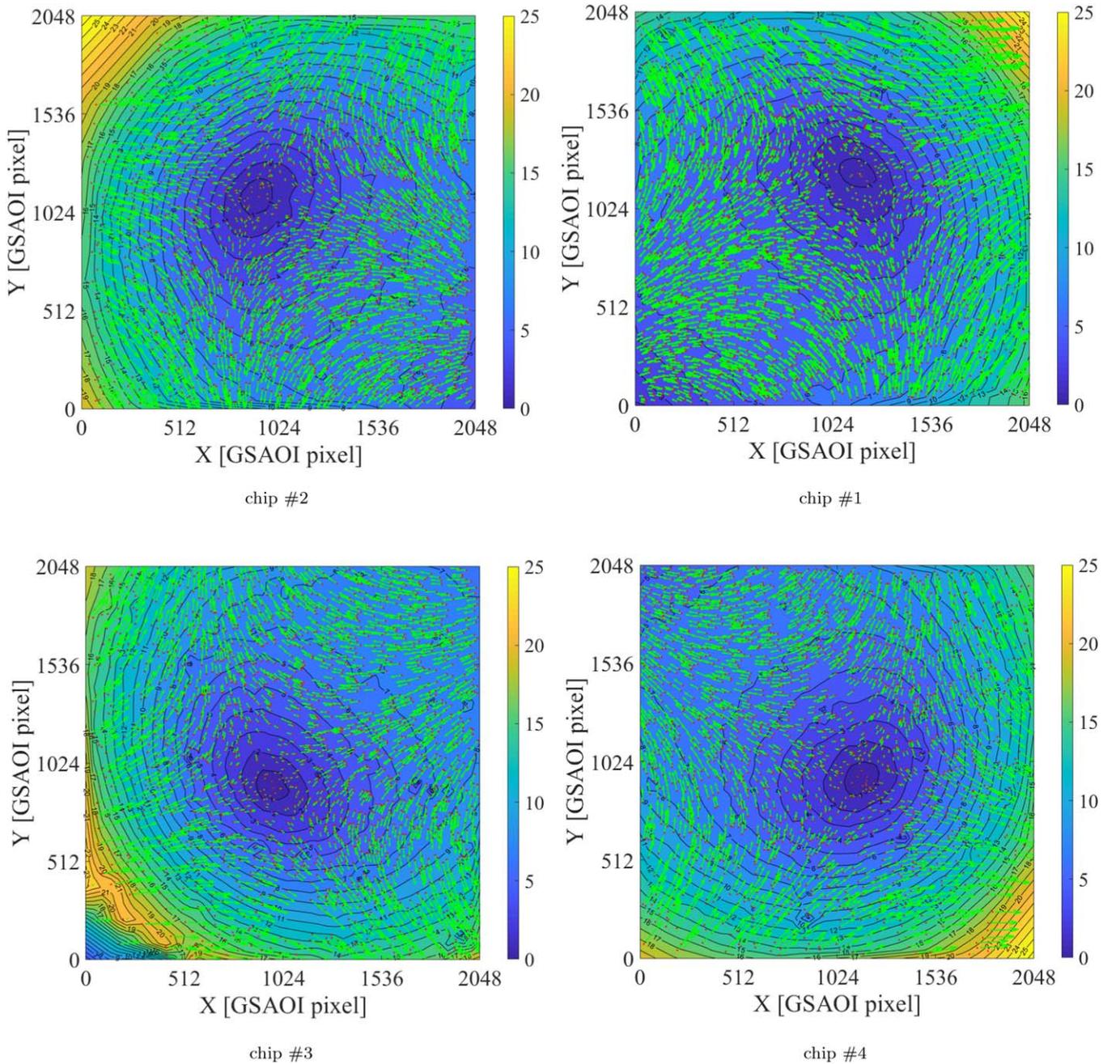

chip #2

chip #1

chip #3

chip #4

**Figure 6.** All four final displacement vector fields (DVFs) for exposure #4. The background color represent the intensity of the displacement between the two catalogs for each point in the field. The green vectors show the direction of the displacement. Each GSAOI pixel is equivalent to 0″.02.

To extract the distortion model from the initial DVF, we have developed a zonal analysis method, specifically an algorithm that looks at the behavior of each vector compared to its neighbors. In contrast to previous approaches, such as Kerber et al. (2016) and Dalessandro et al. (2016), our methodology does not require that we spatially bin any of the data or take averages, as we want to keep the intrinsic spatial resolution of the catalogs to probe the distortion field (which is set by the density of matched star pairs). It works on any matched catalog of objects. In essence, if the size and direction of the vector is consistent with its neighbors, the vector is kept, and if not, it is removed. To make this decision for each vector, the algorithm considers multiple features. These features are calculated based on the analysis of the histogram of differences between the DVF and each individual vector in its local neighborhood. This is done iteratively: the *first loop* cycle starts with the first DVF and scans the field repeatedly until the number of remaining stars and the rms of the vector field reach an asymptote. Figure 4 shows a sample of the DVF before and after the first loop cycle. In the case of the DVF before the first loop cycle, there are clearly many mismatched stars that have large vectors pointing in random directions, and which contribute to the high spatial frequencies in the map. Clearly, the high-spatial-frequency behavior of the DVF is removed and only low-spatial modes remain after application of the first loop cycle.

The similarity transformation (that accounts for the scale–shift–rotation of the two fields) is recalculated at the completion of the first loop cycle, based now on only those pairs of stars that remain in the distortion map. The whole process then repeats in the *second loop* cycle. Like the first loop, the second loop cycle continues until the rms of the DVF and the number of participating star pairs reach an asymptote.

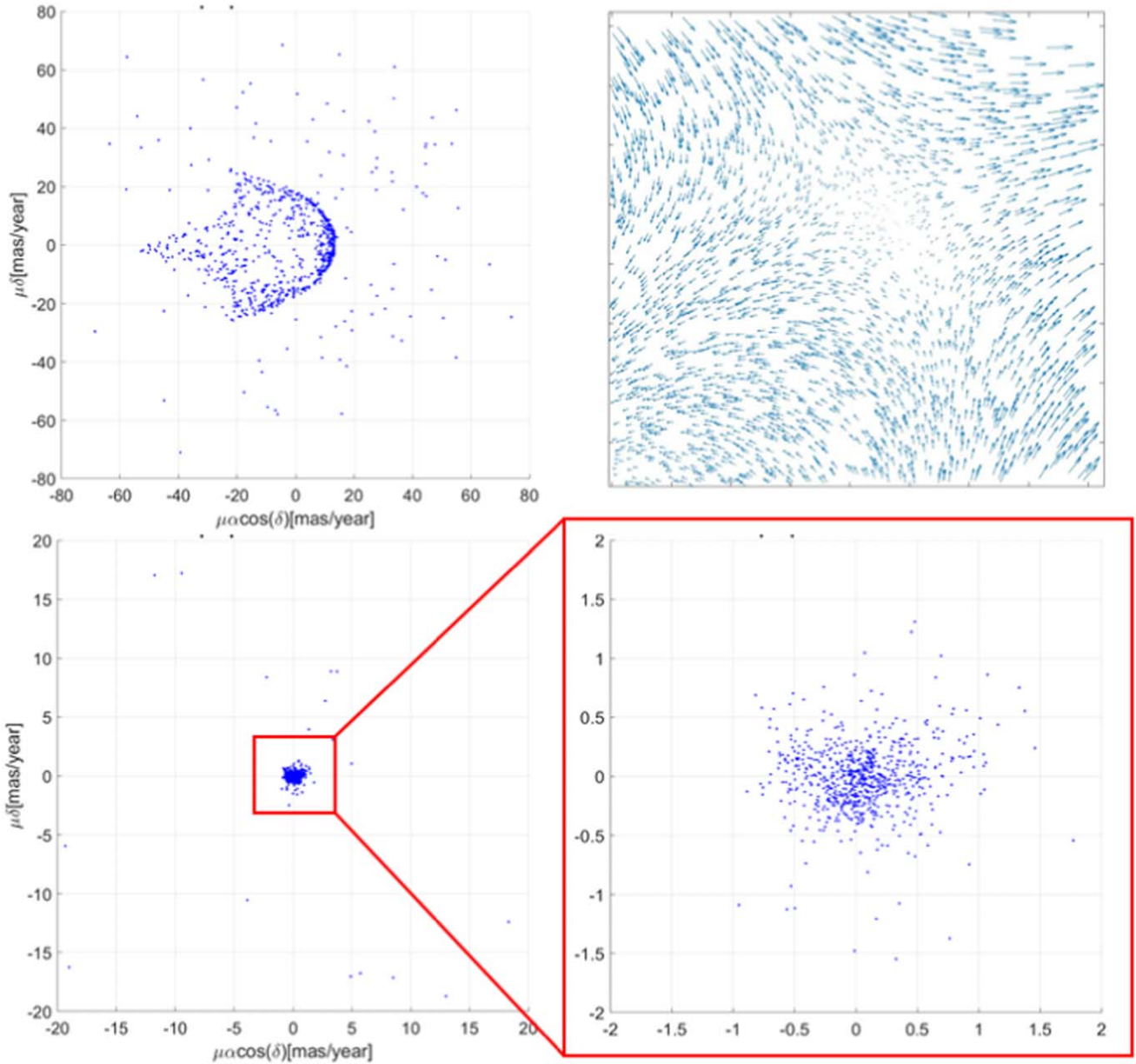

**Figure 7.** Top left: proper motion VPD for NGC 6723 obtained by simple matching of stars between the GeMS and HST/ACS catalogs before the distortion compensation procedure. Top right: differential distortion map for the first exposure for chip 4. Bottom row: proper motion distribution after applying the distortion correction (inverse of the top left panel).

At the end of the second loop cycle, the information remaining in the DVF is the basis for our distortion model. For the example shown in Figure 4 , there are 3251 star pairs with an rms displacement of 167 mas prior to applying the first loop cycle. After the first loop cycle, 2502 star pairs remain with an rms displacement of 135 mas. Figure 5 shows the final result after the second loop cycle, where there are now 2581 star pairs with an rms displacement of 130 mas. Although this figure and the bottom right panel of Figure 4 look very similar, there are subtle differences between the two, as the second loop cycle acts like a fine adjustment to the matching. In this specific example, 79 more star pairs are recovered by the end of the second loop cycle compared to the first, and the rms is reduced by 5 mas. This increase in the number of pairs of stars and the reduction in the rms are due to the fact that the similarity transformation between the two fields is recalculated at the start of the second loop cycle. The sample results shown in Figure 5 are calculated by five iterations of the second loop cycle, with an average of four iterations in the first loop cycle. The same result for all the chips in exposure number 4 (see Table 1 for exposure reference numbers) is presented in Figure 6.

### 3.5. The Final Match between Catalogs

Our GSAOI data set consists of multiple dithered exposures. The dither pattern covers the gap between the detector chips and increases the precision of our photometric and astrometric measurements. Our analysis thus far has been performed on each chip and each exposure individually, for a total of 32 DVFs. To combine data sets, it is easy to transform all exposures to the same frame of reference and to match stars in close proximity. Here, we use all stars in our GSAOI data set. However, given the different epochs of the HST and GSAOI data, this could result in incorrect matches for high-proper-motion stars.

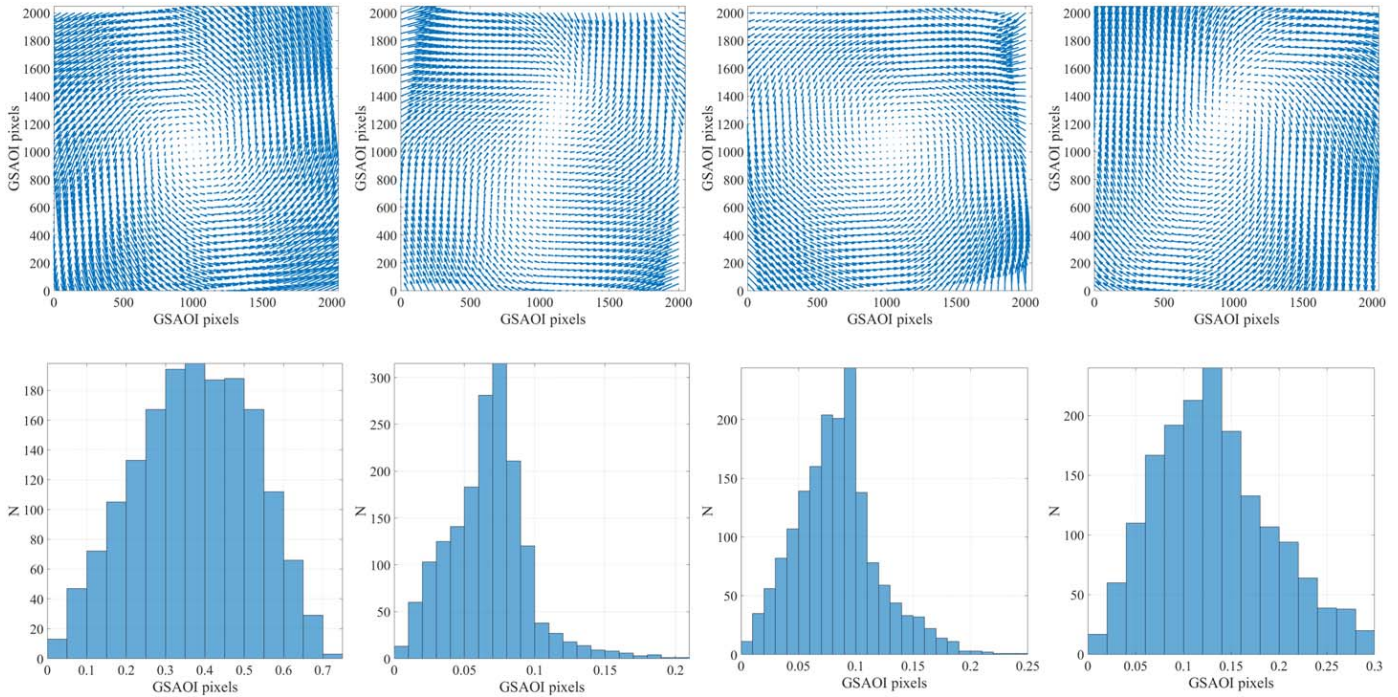

**Figure 8.** Four instances of the measured time variability in the relative distortion maps for chip #4. These diagrams represent the difference of the differential distortion map for exposures 2, 3, 6, and 7 (left to right; see Table 1 for exposure reference numbers) and the first exposure. All these exposures are taken within a period of 15 minutes. The top panel shows the vector field, and the bottom panel shows the histogram of the intensity of the magnitude of the distortion. Each GSAOI pixel projects to 20 $\mu$as on sky. Note that the intensity of the time-variable component is an order of magnitude smaller than the static distortion map shown in Figure 11.

To consider this issue, we first use the nearest-neighbor match only for star pairs that were already confirmed as matching pairs in the second loop cycle of the zonal analysis in the previous subsection. These pairs are used to form an initial CMD. To match the rest of the stars, including high-proper-motion stars, we use this initial CMD in combination with a nearest-neighborhood algorithm. In particular, we ensure that the colors and magnitudes of the matched stars are consistent given the CMD. Any star that cannot be matched with a candidate from the catalog is identified as noise and removed from the process.

## 4. Distortion Maps

The previous section describes in detail how we match different catalogs and use the resulting DVFs to create distortion maps. This information is essential to correct the proper motion measurements we are trying to make, and we describe that process in Section 4.1. Our method differs markedly from previous approaches in the literature (e.g., Dalessandro et al. 2016; Kerber et al. 2016). In particular, we use all photometrically identified sources in our analysis, regardless of magnitude (i.e., we do not limit ourselves to only analyzing the brightest stars), and we do not bin or average our data. In this way, we use data with the highest spatial resolution possible (set by the density of matched star pairs) on which to fit the distortion model, making maximal use of our data. Our only assumption is that the distortion model be continuous and dominated by low-spatial frequencies. The methodology described in Section 3.4 can be applied to any two catalogs that have been matched; however, how the resulting distortion field is subsequently interpreted needs careful consideration and depends on the catalogs that have been matched.

In the case of the GSAOI and HST/ACS catalogs that form the basis of our analysis so far, the distortion map is actually a *differential* distortion map. That is, it provides a map that contains the difference in position caused by the *relative* distortion that exists between the two instrument systems: it includes distortion from the entire Gemini/GeMS/GSAOI system (i.e., atmosphere, telescope, and instrumentation) as well as any distortions that remain present in the HST/ACS instrument system (telescope plus instrument). This means that each single differential map does not give any direct information about the astrometric performance of each system independently, but rather they reveal the entangled displacement effect caused by the distortion fields of both instruments.

In this section, we discuss how we use distortion models to enable our science and, uniquely, to understand the behavior of our observatory system. In Section 4.1, we first discuss the differential distortion maps, and how we apply these as a correction for proper motion analysis. Section 4.2 then shows how we can use these maps to monitor changes in the observatory system with time. In Section 4.3, we demonstrate how it is possible to extract the distortion map due *only to* GSAOI using the data in hand, and we argue that this type of analysis can be used as an important diagnostic tool to test, measure, and monitor the system performance.

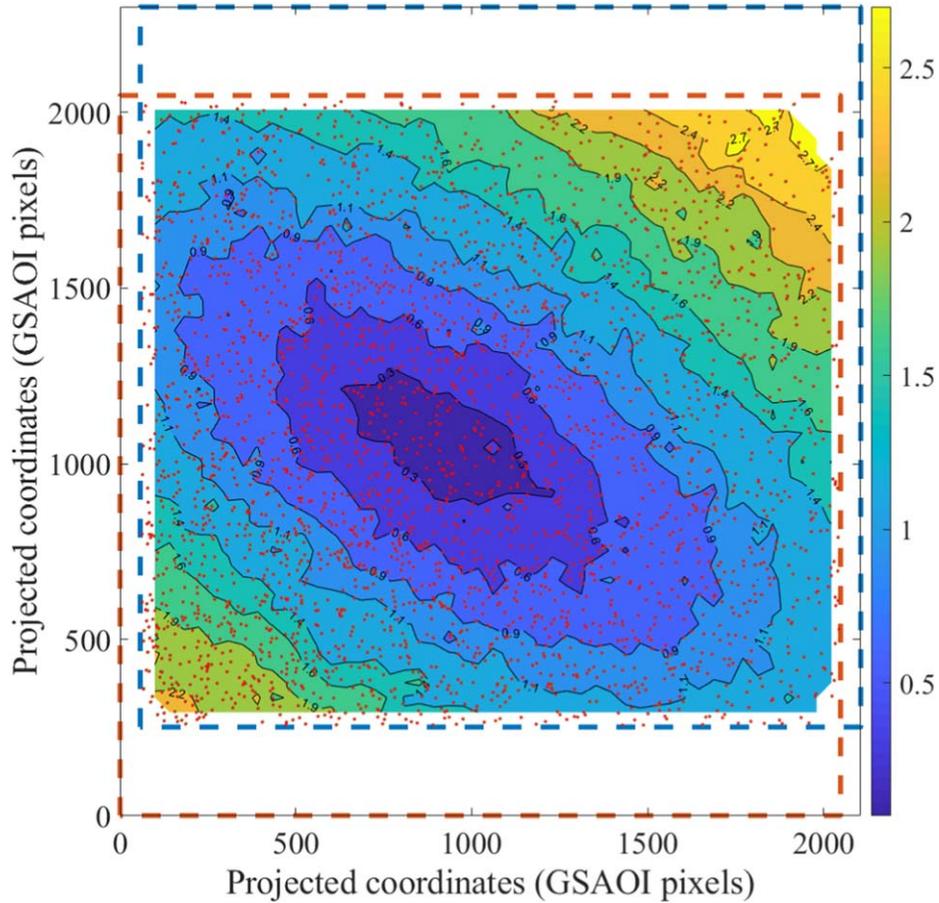

**Figure 9.** The relative DVF between exposures 4 and 5 for chip 4 of GSAOI. The dashed orange and blue squares indicate the relative positions of chip 4 in exposures 4 and 5, respectively. The color bar represents the displacement of star pairs in units of GSAOI pixels (20 mas). The obvious shift between the two dashed-line frames is caused by the telescope dither between the two exposures. The overlapped region contains 2850 verified star pairs.

### 4.1. Differential Distortion Map and Distortion Compensation

With the creation of the differential distortion map, we can model and subtract the effect of distortion on our astrometric measurements. We fit degree-five polynomials (Massari et al. 2016c) on each dimension of the processed DVF using all the stars pairs that passed the process described above. This allows us to create the distortion model for any particular chip and exposure. There are two reasons to fit a model on the DVF to form the distortion model:

1. The DVF includes the distortion information only for the points for which a pair of stars that survive previous steps exists. To interpolate at any point across the field of view requires some type of model;

2. Because of the discreet nature of the DVF (the fact that there are only pairs of stars at some and not all positions in the field), a small component of random relative proper motion can pass through the multiple-iteration process. This component originates from the internal motion of stars relative to the rest frame of the globular cluster. Fitting a continuous low-spatial-frequency model on the DVF therefore acts as a way to fit over these (small) residuals.

Armed with the fitted distortion model, it is trivial to calculate the distortion-correction vector field by inverting the distortion model functions. Figure 7 provide a visual representation of the effect of the distortion compensation on proper motion measurements represented as a vector point diagram (VPD). The VPD is a diagram in proper motion space, where each point represents the proper motion of a star. The top left panel shows the "raw" VPD for stars on chip #4, exposure 1 of GSAOI. Note the scale of the proper motions is of the order of tens of mas yr$^{-1}$ and is irregular. The top right panel is the differential distortion maps for this chip, corresponding to the first exposure. The inverse polynomial fit to this map is applied on the DVF, and the results are shown in the lower panels. Here the distribution of points is much more regular and has a scale of less than 1 mas yr$^{-1}$, i.e., the relative distortion between Gemini/GSAOI and HST/ACS is an order of magnitude larger than the science measurements that we seek to make.

### 4.2. Time-variable Distortion Maps

In addition to distortion compensation for astrometric analysis, differential distortion maps such as those discussed above have a great potential for monitoring instrument performance over time with the aim of improving astrometric performance.

We illustrate the utility of this idea by measuring the time-varying component of the distortion field over the course of our observations. We perform this analysis by fixing the similarity transformation between the two catalogs for all eight GSAOI exposures and look at the *difference* of the differential distortion map for each exposure relative to the first one. The result is presented in Figure 8 for a few of the exposures. The top row shows the vector fields, and the bottom row shows the histograms of the magnitude of the vectors in the top panels. Typically, the magnitude of the temporal distortion residual is of the order of 0.1 pixel (2 mas), although for exposure 2 the magnitudes of the residuals are larger. This is in broad agreement with the temporal distortion residual value found by Massari et al. (2016b).

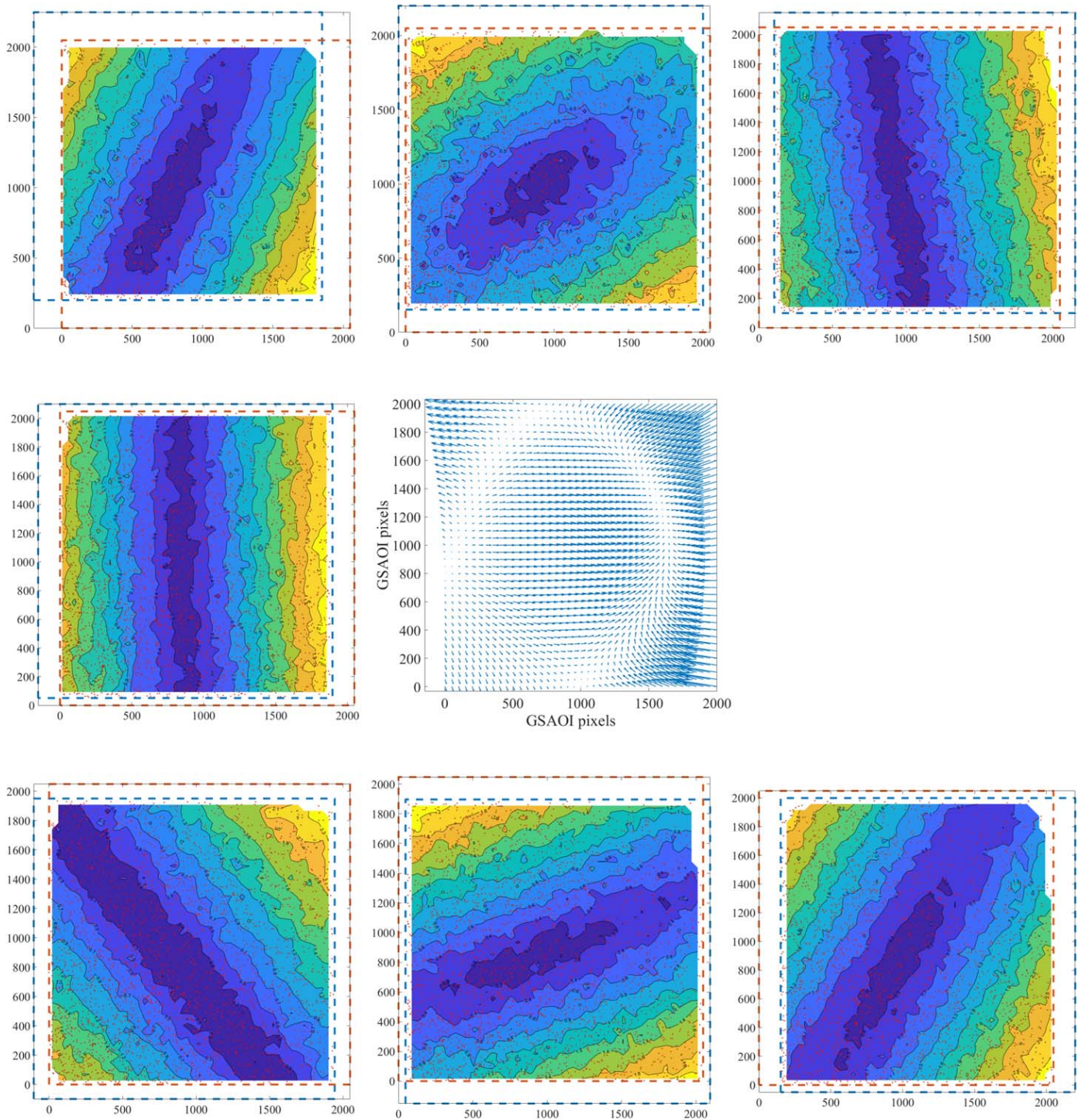

**Figure 10.** Each panel around the outside represents the differential distortion field of chip #4, relative to exposure #5. Each DVF is approximately positioned relative to the central panel by its dithered field location. Combining these maps allows us to construct the static distortion map (central panel). If the static distortion map is displaced relative to itself and subtracted, it would create the differential distortion maps that are observed.

Examination of the vector fields in the top row of panels in Figure 8 reveals that the majority of this time-varying distortion component over the course of our observations consists of rotational modes. The effect is at the subpixel level and points quite convincingly to the field rotator as the dominant sources of this time-varying distortion, an interpretation that is consistent with an independent study by Riechert et al. (2018). Clearly, examination of the differences between the differential distortion maps of an instrument system taken at different times can provide a potentially very powerful diagnostic of time-varying distortions, enabling a better understanding of the scale and sources of systematics in astrometry that will otherwise dominate astrometric error budgets.

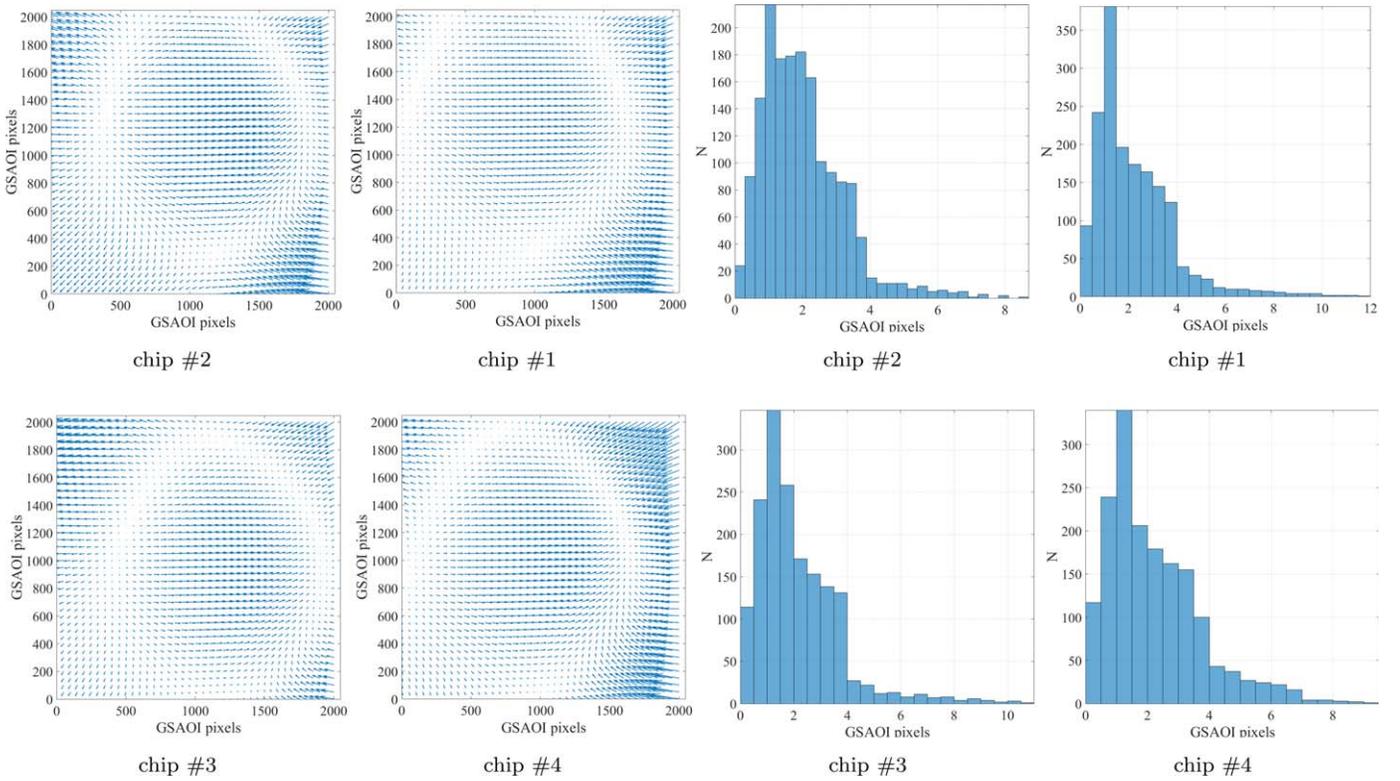

**Figure 11.** Static distortion map and displacement histograms for all four chips of GSAOI. The vector field average of each chip is removed.

### 4.3. On-sky Measurement of Static Distortion Maps

Using the same concept as described for the creation of the differential distortion map between the GSAOI data and HST/ACS, it is also possible to create a "static" distortion map, that is, a map showing the common distortions between all of our Gemini/GeMS/GSAOI observations. In contrast with the differential distortion map, the static distortion map only contains distortion components that are fixed relative to the field of view of the instrument and are preserved, quasi-static, during the on-sky dithering process.

To make the "static distortion" measurement, we only use the dithered GSAOI observations. In this way, our procedure becomes similar to the well-known autocalibration methods used by Anderson & King (2003), Bellini et al. (2011a), and others, albeit formulated in a different way as we utilize the knowledge of the dither pattern in the process of estimating some low-order distortion modes. We follow the same procedure described earlier but now applied only to the overlapping regions of each pair of dithered exposures. We also use a "shift-only" transformation model instead of a similarity transformation, as we only expect a shift due to the dithering process, and any rotation or scale in this analysis are considered as components of the instrument distortion. Figure 9 shows the differential distortion field for the overlapping sections of chip number 4 for exposures 4 and 5.

A differential distortion field like Figure 9 can be calculated for each pair of dithered exposures. Using each differential distortion field, we then solve a nonlinear differential vector field problem, where the goal is to reconstruct a static distortion field that can provide all the differential distortion fields over the specific dithering vector for each pair of exposures.

The outer seven panels of Figure 10 show the input differential distortion maps for chip 4, where there are seven differential distortion maps because there were eight individual exposures. The central panel is the reconstructed static distortion field, that is, the elements of the distortion field that are common to all the exposures. If this derived static distortion field is displaced along the dithering vector and subtracted relative to itself, it would produce each of the seven observed differential distortion maps shown in Figure 10. The complete static distortion map and displacement histograms for each chip are shown in Figure 11.

Given the duration of our exposures, very rapidly changing effects with temporal frequencies of the order of seconds and minutes will be averaged out and will not be a significant contributor to the static distortion maps visible in Figure 11. Thus, these maps are a suitable tool to study the effects of instrument flexure, optical design, some aspects of MCAO system performance like residual modes between two DMs, and other parameters affecting instrument performance. This procedure has the same shortcomings as the autocalibration method used by Anderson & King (2003), Bellini et al. (2011a), and others. In particular, the input data that we have used are not equally sensitive to all the spatial distortion modes that might be of relevance, as this is set by the dither pattern of the input images, and the spatial scale of the offsets that each pair of images probes. Ideal observations for derivation of the static distortion map of GeMS/GSAOI would include a much more extensive dither pattern, possibly including rotational dithers.

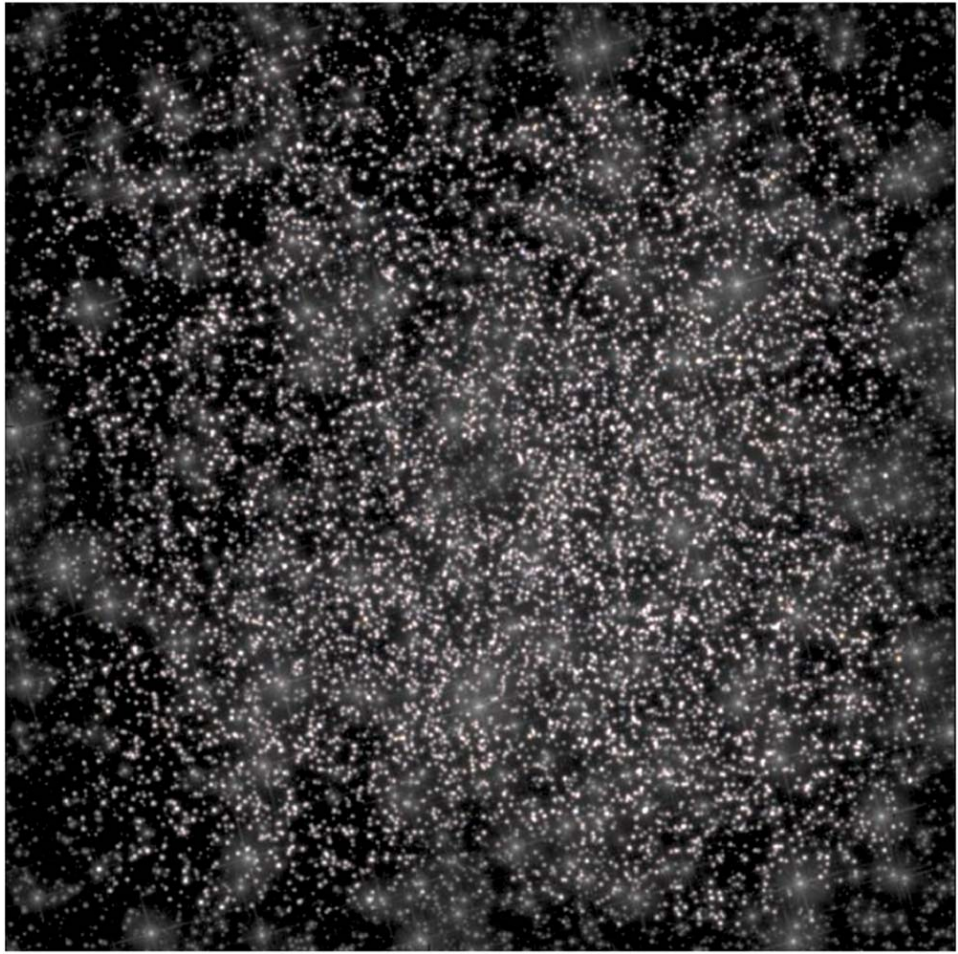

**Figure 12.** Animation showing the linear extrapolation of measured proper motions over time, in the absence of gravity. The initial image is the HST field of view. Zooming in to the core of the cluster, the image fades to the second-epoch GSAOI-measured position. The color of each star is representative of its measured $F606W - K_S$ color. Each second in this video is equivalent to 30 yr.

(An animation of this figure is available.)

It is revealing to compare our results with Figure 3 of Patti & Fiorentino (2019). These authors derive the expected distortion map of GeMS/GSAOI due solely to its optical design. Despite our unequal sensitivity to all spatial modes, it is striking that, although our map applies to the entire observing system (i.e., including the telescope and the AO system), the general behavior and intensity of our on-sky measurements are in very good agreement with this study. This implies that the optical design of GeMS/GSAOI is the dominant term in the static distortion component for this observing system.

We believe that these techniques described in this section could become essential tools for future generations of astronomical instruments, especially the ELTs, to monitor, diagnose, and understand the actual astrometric performance of these observing systems. This will be especially true when used in conjunction with internal calibration microaperture grids (Rodeghiero et al. 2016; Crane et al. 2018; Service et al. 2019) as an on-sky verification of measurements. One clear advantage of the methodology presented here is our reliance on on-sky observations, matching closely the setups required for actual science measurements (indeed, these diagnostic tools can be produced from the actual science observations).

## 5. Proper Motion Measurements

### 5.1. Relative Proper Motion Measurements

By combining data from all exposures and chips using the method explained in Section 3, it is possible to calculate the relative proper motion VPD for the cluster. We require that each star is observed a certain number of times across all GSAOI exposures, and this acts as an additional quality control criteria. By increasing this threshold, the precision and reliability of our data increases but at the cost of smaller numbers of stars meeting the cut. Increasing the verification threshold decreases the number of data points in a semilinear trend but increases the measurement precision of each data point. We adopt a verification threshold of three independent detections, which results in a total of ~12,500 data points. We measure the proper motion of each star by calculating its mean position in the distortion-corrected GSAOI exposures and comparing them with the corresponding first-epoch HST positions. The uncertainty in the proper motion of each star is estimated as the random error in the mean of the multiple measurements for GSAOI, combined with the random error in the HST positions described in Section 2.1. As discussed in the previous section, the displacement caused by the combined systematic uncertainties of GSAOI and HST has been removed by application of the *differential* distortion correction, and we will later examine the scale of any remaining systematic effects not accounted for by this correction. The VPD for NGC 6723 analyzed in this way is shown in Figure 13.

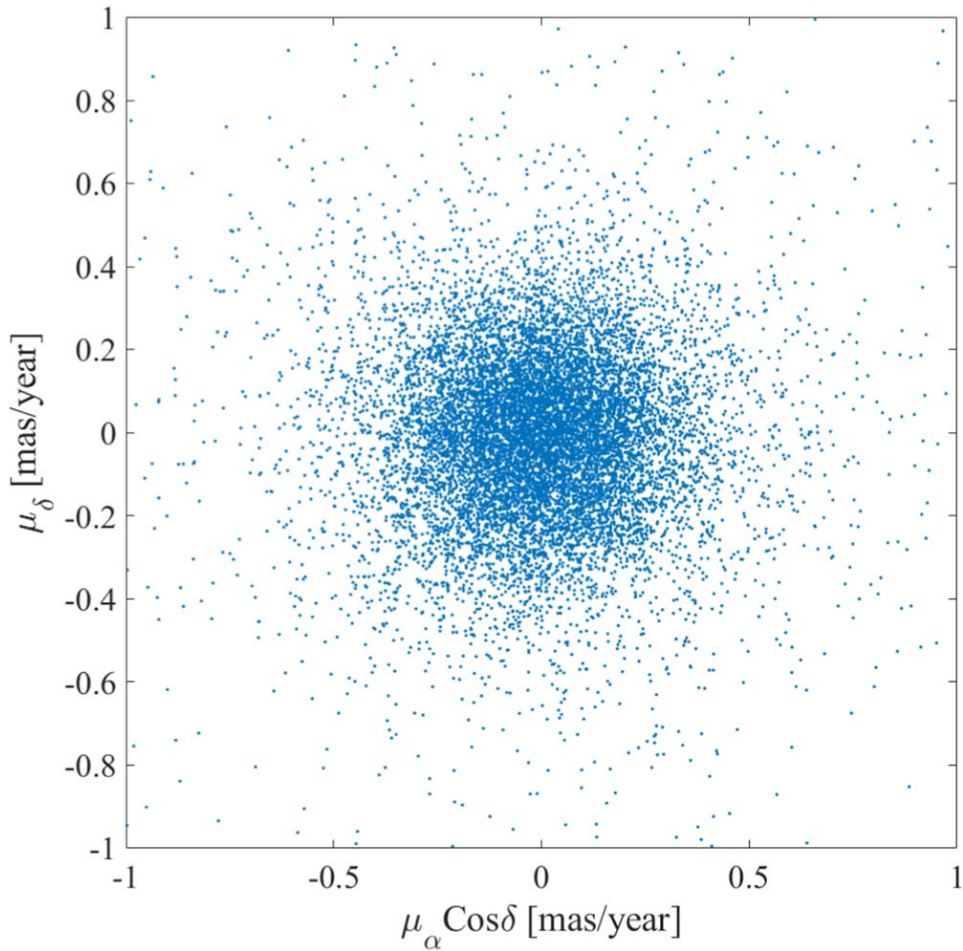

**Figure 13.** Proper motion diagram for NGC 6723. Each point in this diagram represents the proper motion of one star. This representation is for the verification threshold of three, containing ∼12,500 data points.

To summarize all measurements in a nice representation, we produced a short animation (Figure 12). This short video shows the initial HST measurements as a fits file and gradually fades to the more recent GeMS/GSAOI measurements. The color of each star is matched with its $F606W - K_S$ color. The motion of each star is the linear extrapolation of the measured proper motion in this study, in the absence of gravity. The timescale of the video is 30 yr s$^{-1}$. In the last few seconds of the animation a closer field of view is shown for better representation of the faint and slow-moving stars. The hole in the center of the video is because by the dither pattern of GSAOI and crowding effects limiting the number of independent detections in this region.

### 5.2. Systematic Sources of Uncertainty

We check the existence of systematic trends in proper motion with the magnitude and color. Figure 14 shows these diagrams. The red line in these figures represents the average value in each bin. The error bars are the standard in the error of the mean for each bin. No significant trend between the magnitude and proper motion values is evident.

We looked at maps of the residual distortion plots to check if any low-frequency spatial trends remain after our distortion correction. Figures 15 and 16 represent the distortion map for multiple chips for exposure #4. Figure 15 shows the distortion map binned by the $x$ and $y$ position prior to correction for the distortion correction, where each chip has been corrected for distortion. Figure 16 shows the corresponding residual distortion map after correction for the distortion, where now vectors have been enlarged by a factor of 2500. The random behavior of the residual distortion maps shows that the distortion reduction process provides good distortion compensation. It should be noted that the final differential distortion model is measured based on the combination of data for all eight exposures, where each chip has been corrected for distortion individually. Therefore we expect that the error budget for the residual distortion in our final result will be less than what is presented in Figure 16 by a factor of $\sqrt{8}$. We also show the overall dependency of the residual distortion as a function of $x$ and $y$ before and after the correction process. Figures 17 and 18 show the result of this analysis for exposure #4. In line with the 2D maps, we see no evidence for significant spatial dependency in the residual plots for any of the exposures and/or chips.

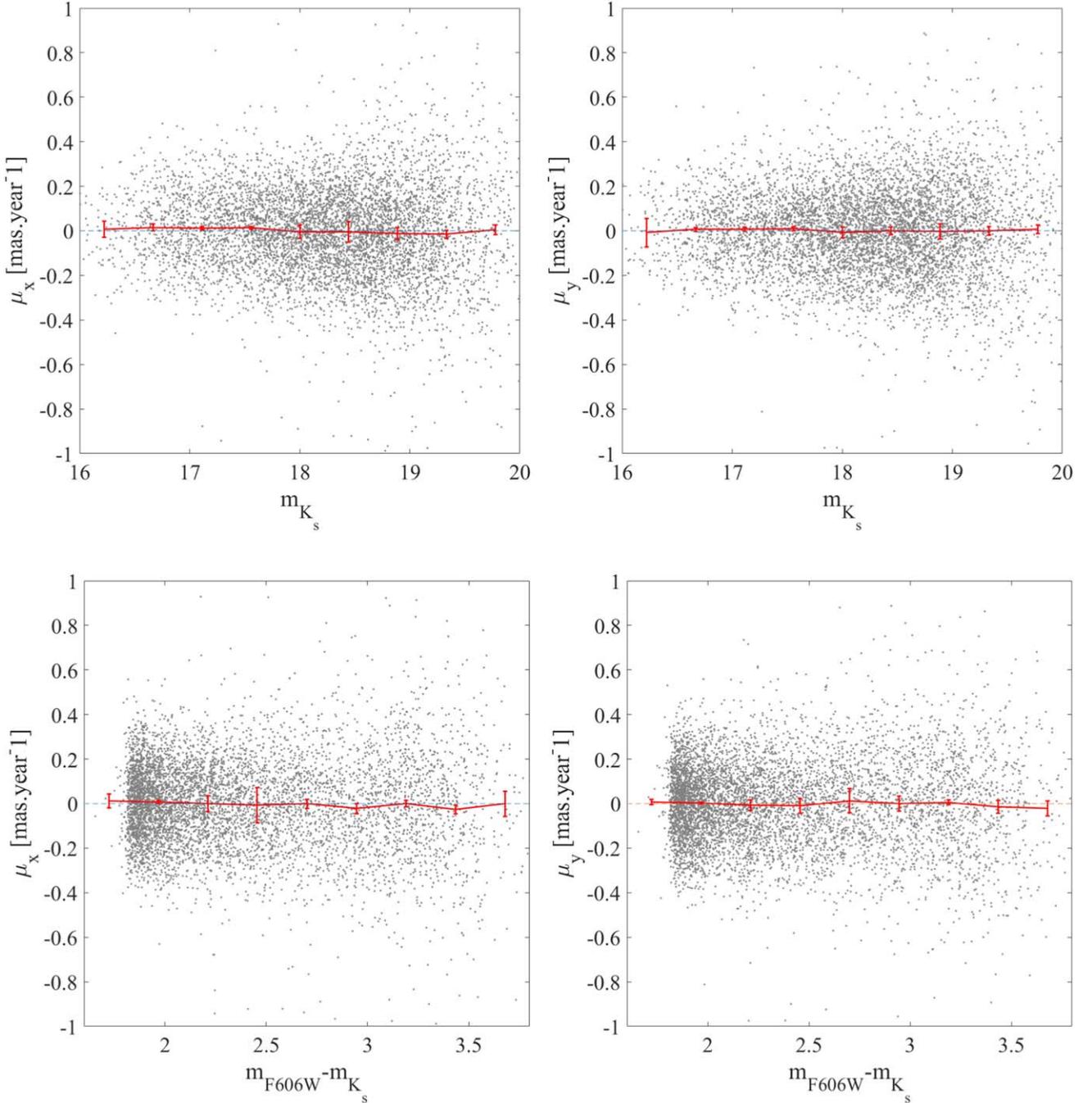

**Figure 14.** Top panel: proper motion components in the x and y directions vs. $K_S$ magnitude. Bottom: proper motion components vs. optical-infrared color. The red line and error bars indicate the mean and the standard error in the mean for each bin, respectively. No significant systematic trend can be seen between the magnitude/color and proper motion.

### 5.3. Tangential Velocity Dispersion Profile

To calculate the dispersion in the proper motions, we adopt a maximum-likelihood technique in order to disentangle the contribution of measurement errors from the intrinsic dispersion of the cluster. We follow the method of Martin & Koelfgen (2007) and others, by seeking to maximize the likelihood function,

$$L(\mu_T, \sigma_T) = \prod_{i=1}^{N} \frac{1}{\sigma_{\text{tot}}} \exp\left[-\frac{1}{2}\left(\frac{\mu_T - \mu_i}{\sigma_{\text{tot}}}\right)^2\right], \quad (1)$$

where $\mu_i$ is the individual proper motion of each star ($\mu^2 = (\mu_{\alpha} \cos \delta)^2 + \mu_{\delta}^2$), $\mu_T$ is the mean proper motion of the entire cluster and is zero by design, and $\sigma_{\text{tot}} = \sqrt{\sigma_T^2 + \sigma_i^2}$ is the sum of the dispersion from measurement errors, $\sigma_i$ (that includes both the GeMS and HST positional uncertainties) and the intrinsic tangential dispersion, $\sigma_T$. Applied to the entire data set, we find that $\sigma_T = 0.148^{+0.002}_{-0.002}$ mas yr$^{-1}$. This corresponds to a tangential velocity dispersion of $\sigma_T = 5.84^{+0.07}_{-0.07}$ km s$^{-1}$. Error bars are random errors only and are small because of the large number of stars involved in the analysis. For calculation of the

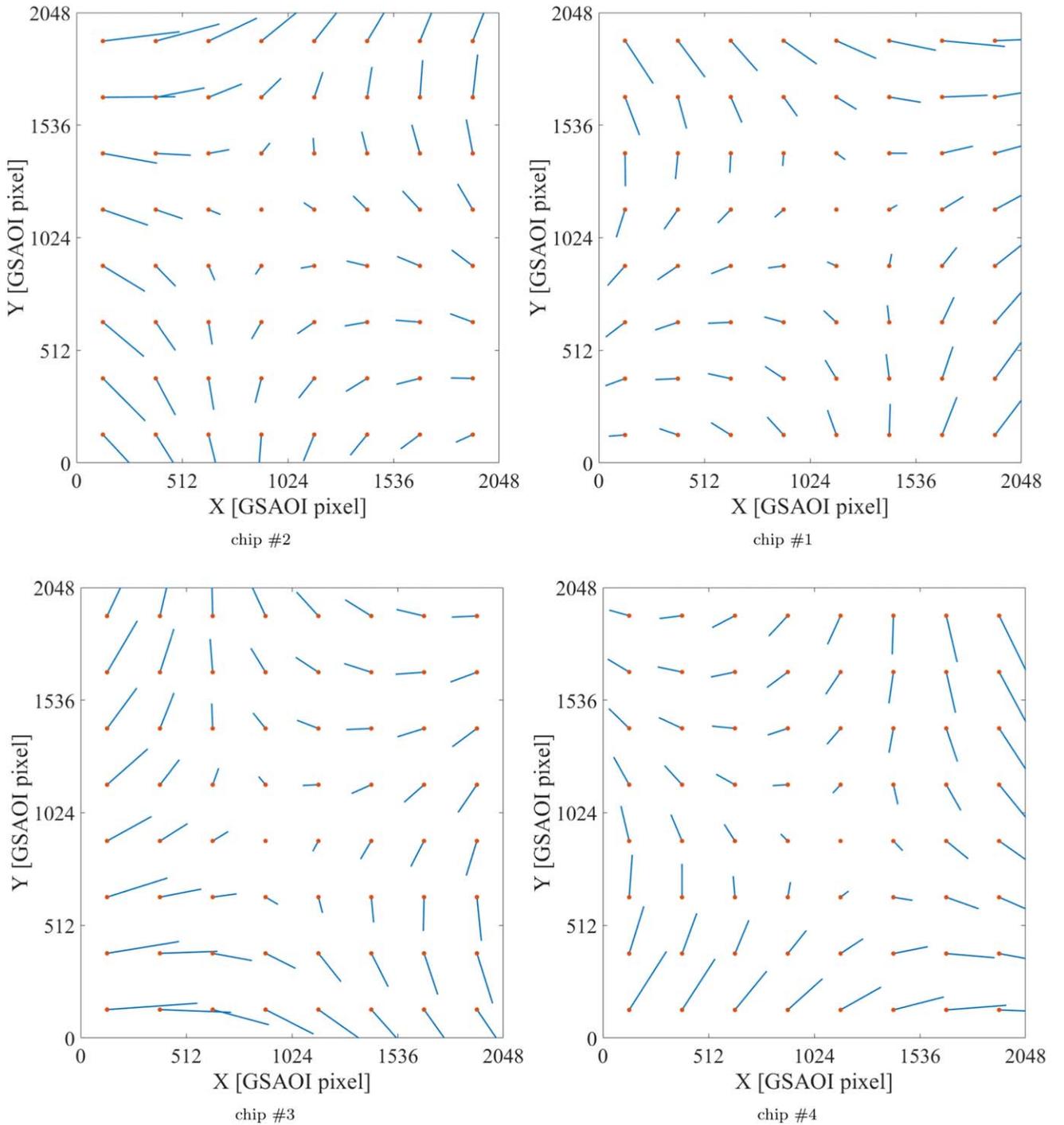

chip #2

chip #1

chip #3

chip #4

**Figure 15.** GSAOI distortion map for exposure #4 before the distortion correction, magnified by a factor of 20.

velocity dispersion, we have assumed a distance to NGC 6723 of $8.3 \pm 0.47$ kpc (Baumgardt et al. 2019).

Additional confirmation of the absence of any significant sources of systematic uncertainties as a function of the magnitude or crowding can be obtained by recalculating the velocity dispersion using only brighter stars, or less crowded stars. The left panel of Figure 19 shows the result of recalculating the velocity dispersion using different magnitude limits, for stars brighter than an $m_{F606W}$ of 20th to fainter than 24th mag. Whatever magnitude cut is used, the result is consistent to well within the $1\sigma$ uncertainties to that obtained using all of the stars.

The right panel of Figure 19 shows the recalculated velocity dispersion using stars in different local environments, as defined by their separation from their nearest neighbor as determined from the HST catalog, in units of the HST-ACS F606W filter PSF half-width-at-half-maximum (HWHM).[9] Here, we expect stars that are more widely separated from their nearest neighbors to be least affected by any residual systematics caused by crowding in the HST data, as discussed in Section 7.1 of Anderson et al. (2008). We go out to 10 HWHM, at which point we have excluded approximately 50% of the stars. While a small trend is present, as expected, the derived velocity dispersion changes by only a small amount. We conclude that the global velocity dispersion and its uncertainties derived above are a good representation of the data, even accounting for residual crowding effects.

---

[9] Nominal HST-ACS F606W filter PSF FWHM is equal to 74 mas or 1.48 pixel (Windhorst et al. 2011).

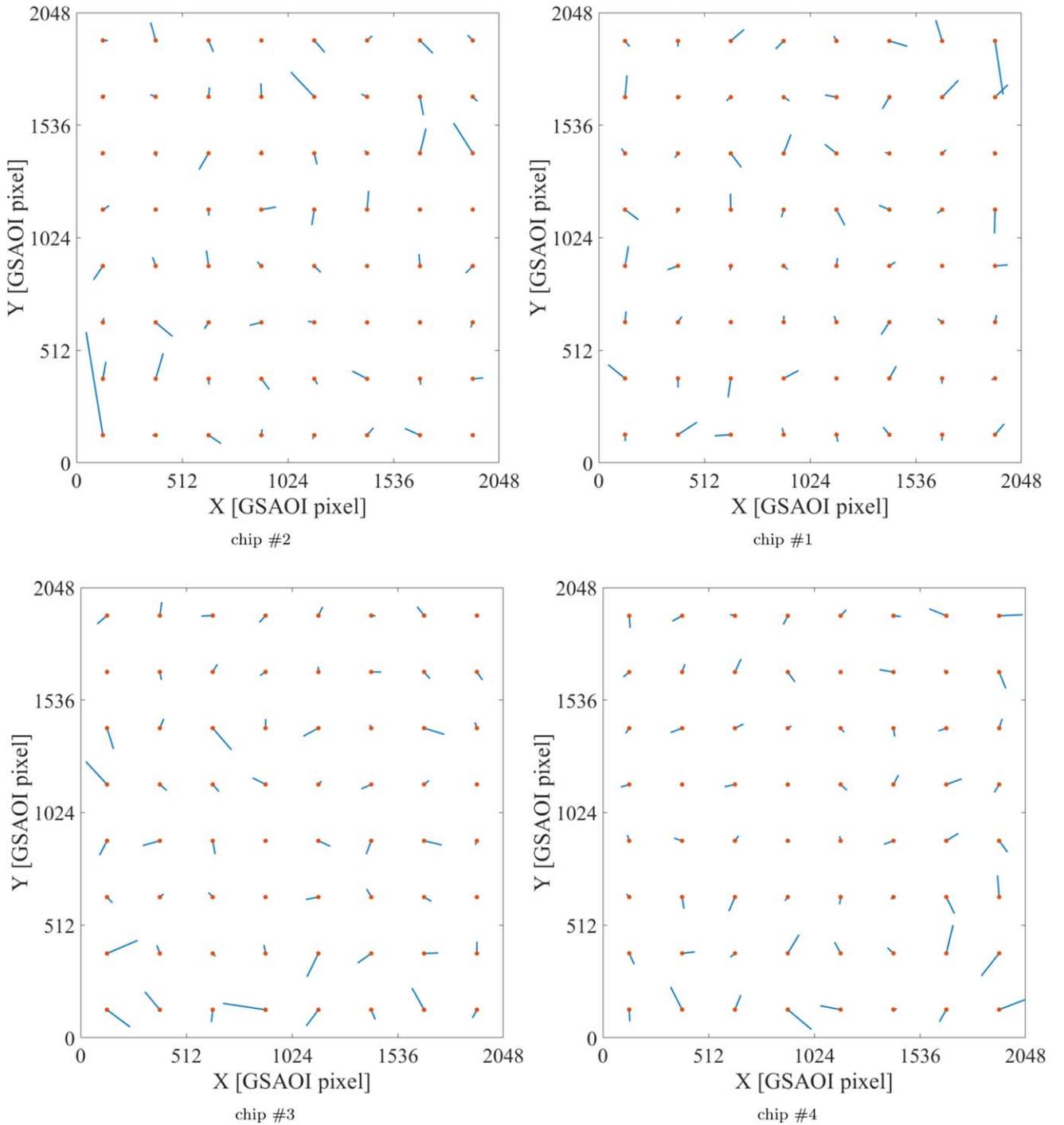

**Figure 16.** Residual distortion map for exposure #4 after the reduction process, magnified by the factor of 2500. No significant systematic trend can be seen in the residual distortion map.

Given the spatial extent of our data, we are also able to repeat this analysis but now examining the tangential velocity dispersion as a function of the radius. To do this, we divided the cluster into concentric regions around the center with equal numbers of stars in each annulus. The centroid of the cluster is adopted from Baumgardt et al. (2019). We measured the velocity dispersion in each annulus using the same MLE algorithm as before. The position of these annuli superimposed on the HST and Gemini data sets are shown in the left panel of Figure 20. In the right panel, the blue data points (dots) show the derived proper motion dispersion profile for NGC 6723. Also shown are the proper motion dispersion measurements from Gaia Data Release (DR) 2, and the line-of-sight velocity measurements, both from Baumgardt et al. (2019).[10] The values of the data points that contribute to the right panel of Figure 20 are given in Table 3. In this table, $R_{med}$ is the median distance of stars to the center of the GC in each annulus alongside the inner ($R_{in}$) and outer ($R_{out}$) bounds of the annulus, $\sigma$ is the 1D equivalent proper motion dispersion, and $V$ is the corresponding dispersion in km/s.

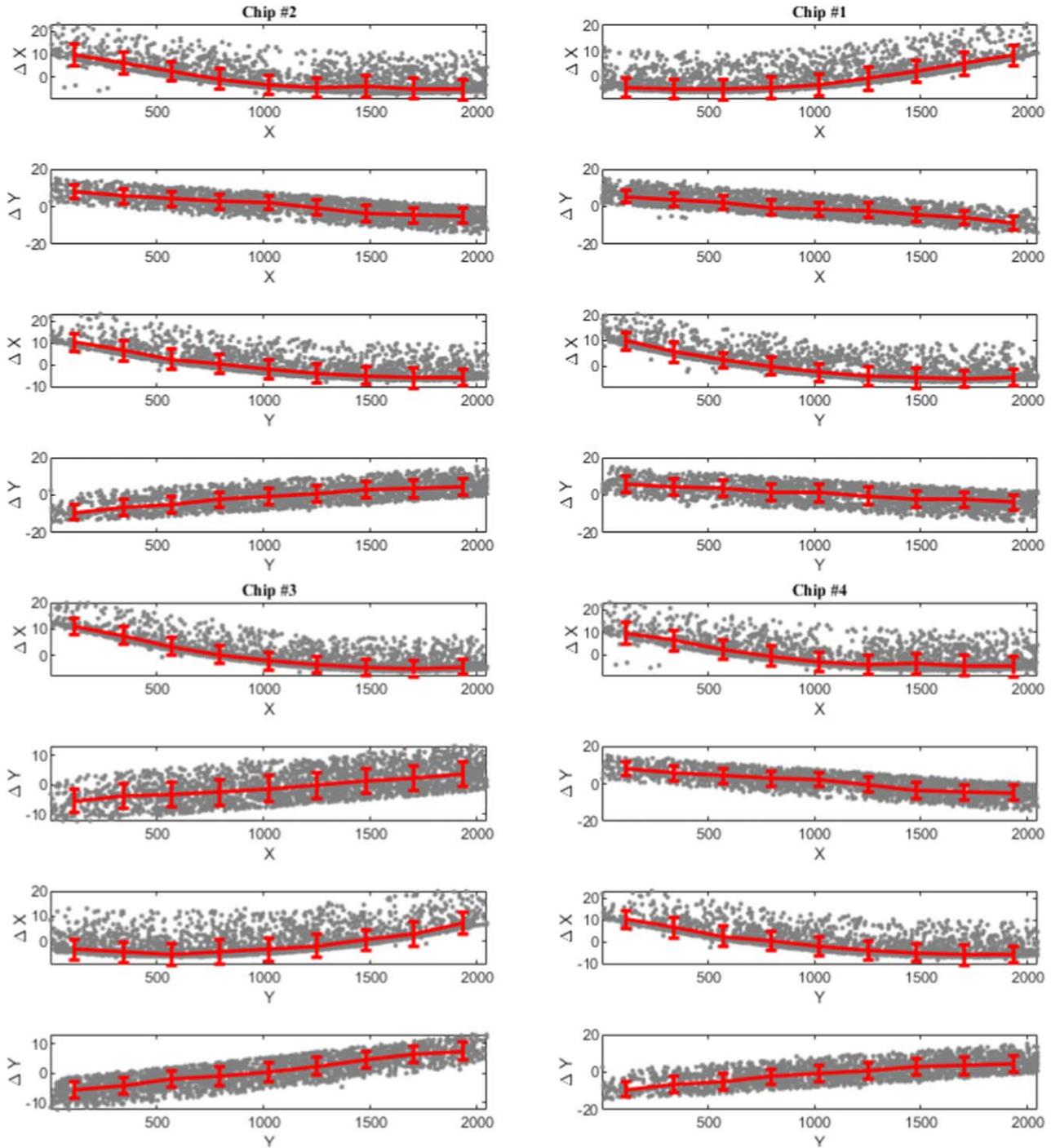

**Figure 17.** Spatial dependency of the distortion before the distortion compensation process for all chips of exposure #4. The unit of the $Y$-axis is GSAOI pixels ($0''02$).

The GSAOI data set is able to probe to considerably smaller radius than either the Gaia or radial velocity data sets, due to the superior image quality and light gathering ability of GSAOI/Gemini telescope. Further, the uncertainties are comparable to, or better than, either of the other two data sets. Finally, we note that the absolute values we obtain are very comparable to the proper motion measurements from Gaia at larger radius and the radial velocity measurements. We conclude that GeMS is able to provide very precise velocity measurements for NGC 6723 to relatively small radii from the center of the cluster.

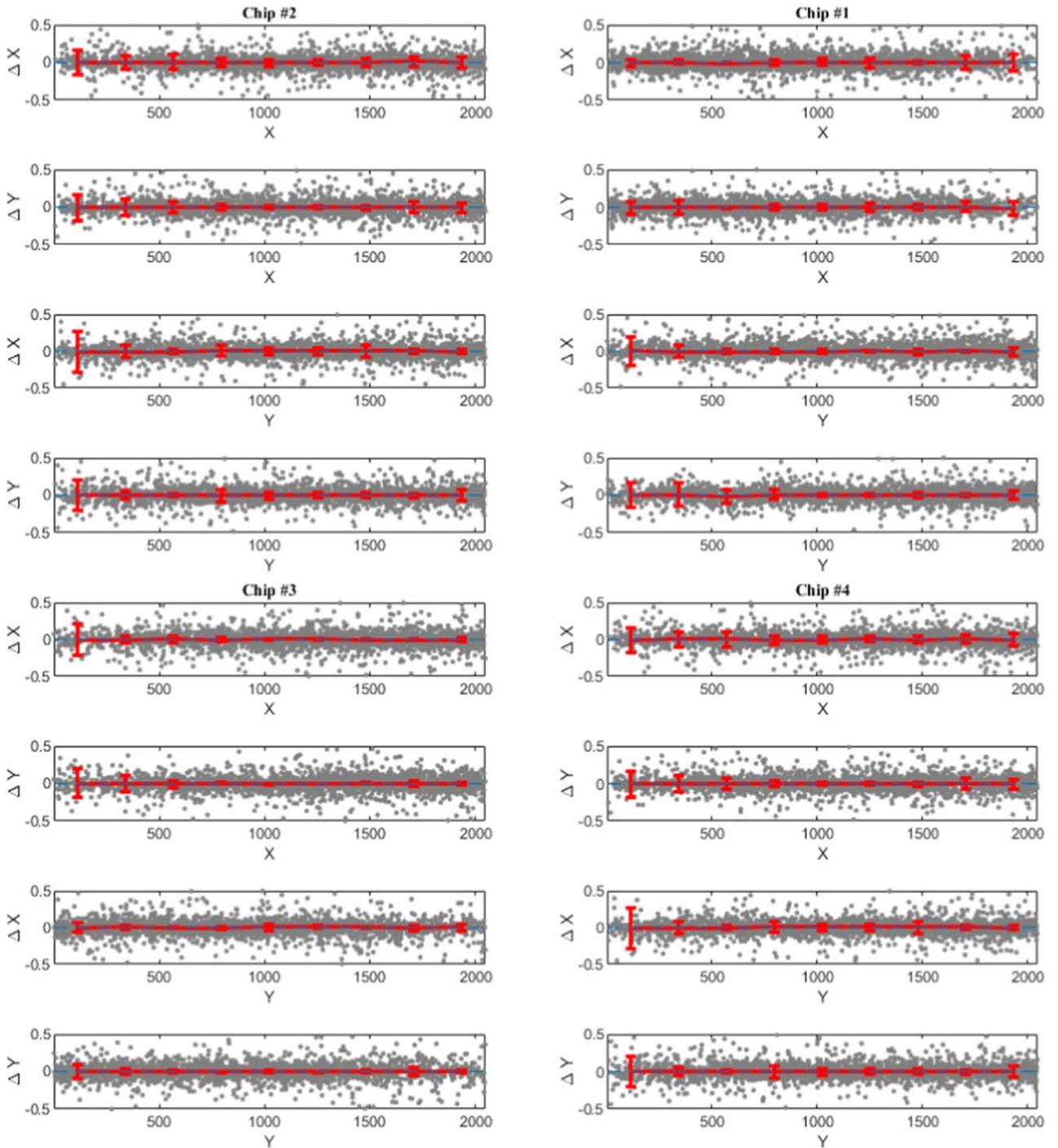

**Figure 18.** Spatial dependency of the residual distortion after the distortion compensation process for all chips of exposure #4. The unit of the *Y*-axis is GSAOI pixels (0″.02).

### 5.4. Comparison to Previous Work

We compare the proper motion precision of our measurement to our group's earlier study using NGC 6681 by Massari et al. (2016b), i.e., using data from the same program and with a very similar time baseline (6.9 yr compared to 6.75 yr for this work). Figure 21 shows this comparison as a function of the instrumental magnitude. We use the instrumental magnitude here as the astrometric precision is primarily related to the number of photons that the detector receives, rather than the apparent magnitude of stars (clearly the latter is a factor in determining the number of photons received, but it is not the only factor; e.g., observing conditions, exposure times). Each dot in this plot represent the precision of a proper motion measurement for a star in NGC 6723 (this work, blue) and NGC 6681 (yellow). The solid lines are the $3\sigma$-clipped mean of each data set and are color coded in red and purple for NGC 6723 and NGC 6681, respectively. It can be seen that the uncertainty floor for this work is given by the bright stars and is approximately $\sim$45 $\mu$as yr$^{-1}$.

The precision of these two studies is comparable, as should be expected for such similar data. Critically, however, the

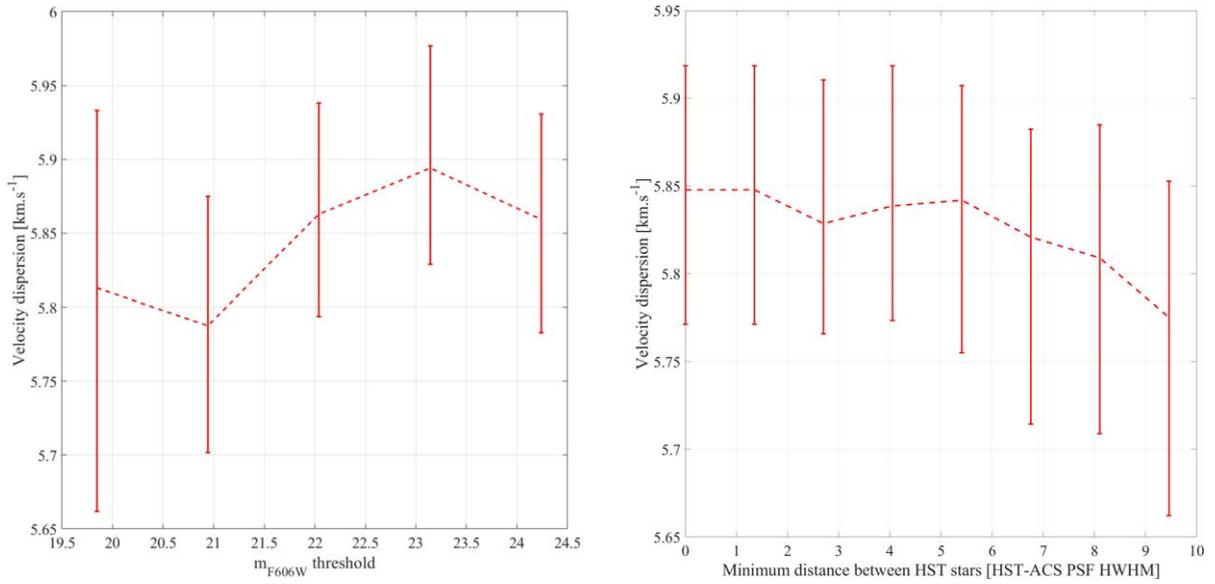

**Figure 19.** Left: global velocity dispersion and its uncertainty as a function of the limiting magnitude as seen in HST. Right: global velocity dispersion and its uncertainty as a function of the local environment of the stars used, defined in terms of their nearest-neighbor separation as determined from the HST data. See text for discussion.

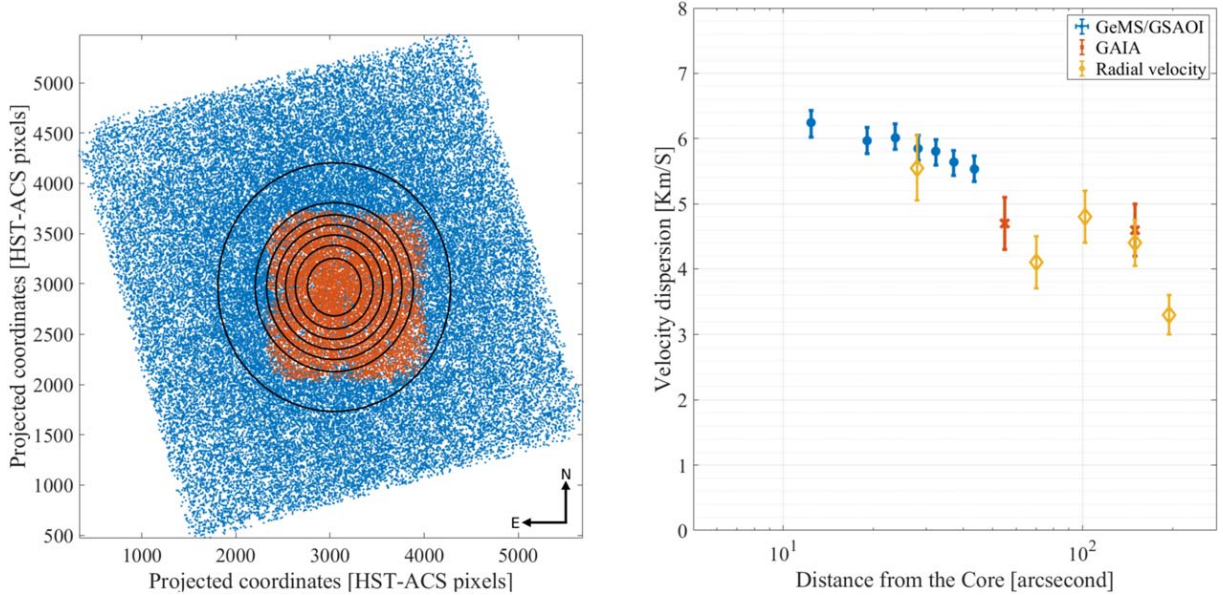

**Figure 20.** Left: the blue and red dots represent stars in the HST/ACS and GeMS/GSAOI catalogs, respectively. The black circles show the boundary of the radial bins, where we have an equal number of stars per annulus. Right: proper motion dispersion profile for NGC 6723. The blue (dot) data points are the 1D equivalent velocity dispersion from this work. Also shown are the velocity dispersion measurements from Gaia DR2 and line-of-sight velocity measurements, both taken from Baumgardt et al. (2019).

**Table 3**
Velocity Dispersion Profile for NGC 6723

| $R_{in}('')|(pc)$ | $R_{med}('')|(pc)$ | $R_{out}('')|(pc)$ | $\sigma_{1D}$ (mas yr$^{-1}$) | $V$(km s$^{-1}$) |
|---|---|---|---|---|
| 0.0–0.0 | 11.6–0.5 | 15.2–0.6 | $0.159^{+0.005}_{-0.006}$ | $6.24^{+0.19}_{-0.22}$ |
| 15.2–0.6 | 18.0–0.7 | 20.2–0.8 | $0.151^{+0.005}_{-0.005}$ | $5.97^{+0.20}_{-0.20}$ |
| 20.2–0.8 | 22.3–0.9 | 24.1–1.0 | $0.153^{+0.005}_{-0.004}$ | $6.01^{+0.22}_{-0.17}$ |
| 24.1–1.0 | 26.3–1.1 | 27.9–1.1 | $0.148^{+0.005}_{-0.005}$ | $5.85^{+0.20}_{-0.18}$ |
| 27.9–1.1 | 30.0–1.2 | 31.7–1.3 | $0.147^{+0.005}_{-0.005}$ | $5.80^{+0.18}_{-0.21}$ |
| 31.7–1.3 | 33.7–1.4 | 36.3–1.5 | $0.143^{+0.004}_{-0.005}$ | $5.64^{+0.18}_{-0.20}$ |
| 36.3–1.5 | 38.0–1.5 | 58.5–2.4 | $0.140^{+0.005}_{-0.005}$ | $5.53^{+0.20}_{-0.19}$ |

methodology presented in this work does not require any additional information or assumptions about the HST/ACS data other than the first-epoch *relative* positions of the stars. In contrast, Massari et al. (2016b) first used the HST–HST proper motion data to propagate the first-epoch HST positions to the second-epoch GeMS data and subsequently remove the distortion field. The methodology presented in this paper is more flexible in being able to derive proper motions from GeMS data that do not require extensive auxiliary data other than (relative) first-epoch positions, and those data are used to construct a DVF that probes the cluster and instrument at the highest possible spatial frequencies. This is a necessary step in

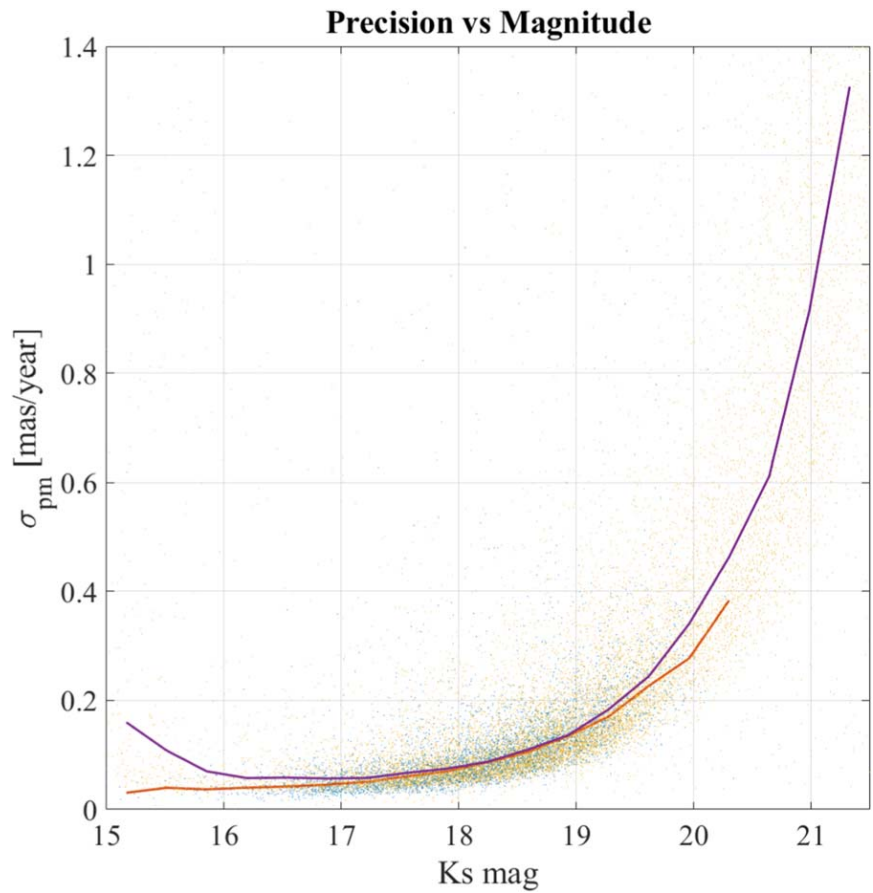

**Figure 21.** Comparison of astrometric precision as a function of the instrumental magnitude for NGC 6723 (this work, blue dots) and NGC 6681 (Massari et al. 2016b; yellow dots). The solid lines are the $3\sigma$-clipped mean of each data set and are color coded in red and purple for NGC 6723 and NGC 6681, respectively. The precision of these two studies is comparable, as should be expected for such similar data. Critically, however, the methodology presented in this work does not require any additional information or assumptions about the HST/ACS data other than the first-epoch positions of the stars. See text for details.

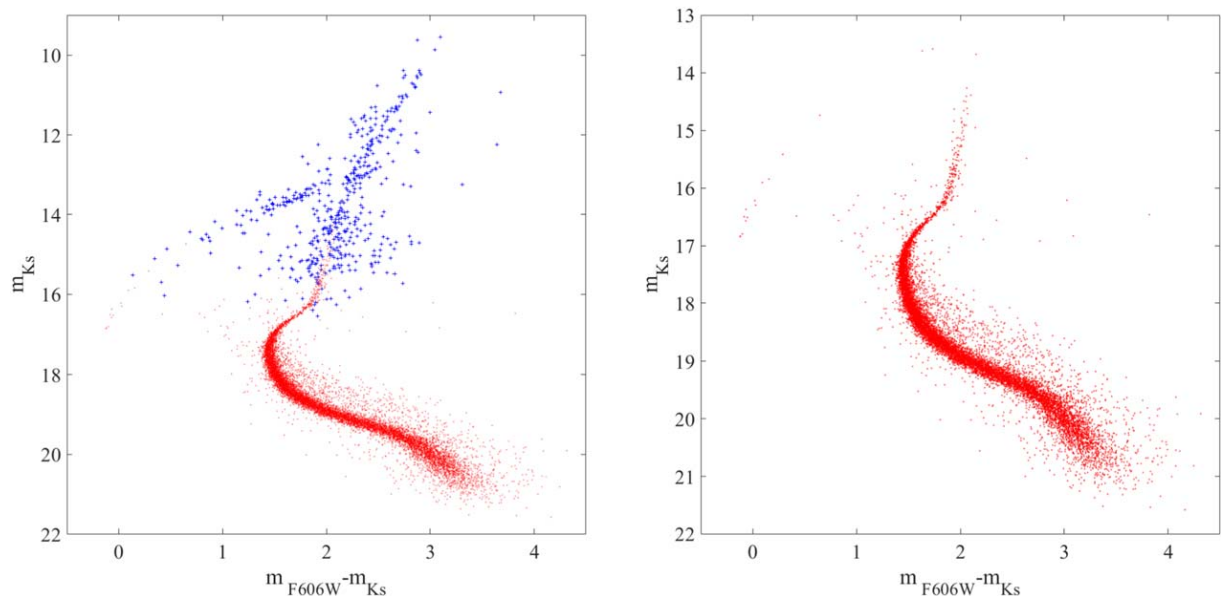

**Figure 22.** The optical–near-infrared CMD for NGC 6723. The red points represent the GSAOI-HST, and the blue points are from the 2MASS–HST catalogs. The continuity of the CMD in the transition region between GSAOI to 2MASS suggests that the photometric calibration is reasonable.

moving toward precision proper motions using only ground-based MCAO observations. We also refer the reader to other relevant works for other globular clusters, for example,

Kerber et al. (2016) and Dalessandro et al. (2016), which also do not require propagating stars to different epochs from which they were observed.

### 5.5. The Optical–Near-infrared Color–Magnitude Diagram

Figure 22 shows the optical–near-infrared CMD for NGC 6723. We note that our technique for identifying matched stars between GSAOI and HST/ACS automatically cleans the CMD of spurious detections or poor matches (i.e., no additional photometric criteria or quality control were needed to create this CMD, and it relies only on the astrometric methods described earlier). As such, the shape and features of the CMD provide strong validation of the matching procedure. In this analysis, we have used only the 160 s Gemini exposures and did not use the short exposures. As such, the final CMD saturates at the approximate level of the horizontal branch. The left panel shows the CMD including ground-based data from 2MASS (used in the photometric calibration discussed earlier), and the right panel shows just the Gemini–HST data. The continuity of the CMD in the transition region between GSAOI to 2MASS demonstrates that the photometric calibration is reasonable. We also note the beautiful extreme horizontal branch stars that are present in the GSAOI data at a color of $m_{F606W} - m_{K_S} \sim 0$ and the main-sequence knee located at $m_{K_S} \sim 20$. A full analysis of the CMD and internal dynamics of NGC 6723 derived from this study will be presented in the next paper in this series.

## 6. Summary and Conclusions

We used GeMS/GSAOI on the Gemini South Telescope to better understand the on-sky astrometric performance of MCAO. We developed a pipeline that can be used to measure and compensate for the relative distortion field between the MCAO observations and any secondary catalog. To validate the performance of our pipeline, we used observations of the core of globular cluster NGC 6723 that ultimately produce a high-quality tangential velocity dispersion profile for the cluster to small radii as well as a deep optical–near-infrared CMD. In order to do this, we had to deal with the adverse effect of the distortion field. We derived the relative distortion maps between the two catalogs by developing a novel method that uses the maximal amount of information to probe the distortion fields at the highest possible spatial resolution (set by the density of stars). We demonstrate how these maps can be used to trace both the time-variable and static components of the distortion. In this way, we are able to show the effect of the Gemini field rotator at the subpixel level, and we can reproduce the distortion caused by the optical design of GeMS/GSAOI. We believe these kinds of measurements will be a valuable tool to diagnose and monitor the telescope/AO system astrometric performance for future generations of large ground-based telescopes.

The methodology presented in this paper does not require any additional information from other sources, other than the *relative* first-epoch positions of the stars being studied. This is a necessary step toward the high-precision MCAO-only proper motions. In the future, this pipeline will be applied to multiepoch MCAO data, an important next step toward enabling future telescopes like TMT and ELT to reach their astrometric potentials.

This paper concentrated on the methodology, the development of the pipeline, and the measurement of various types of distortion maps relating to the on-sky astrometric performance of the system. The next paper in this series will present a detailed stellar population and dynamical analysis of the core of NGC 6723 based on our new measurements.

M.T. would like to thank Dr. Jay Anderson (STSCI), Dr. Benoit Neichel (LAM), Dr. Jean-Francois Sauvage (LAM), and Dr. Lorenzo Busoni (INAF) for their constructive advice regarding this work. He also sincerely appreciates the companionship of his feathery friends, Watermelon the Lovebird and Mango the Sun conure, while writing this paper.

Some data used in this work were obtained from the ACS Globular Cluster Treasury program (PI: Ata Sarajedini, HST Program 10775).

Based on observations obtained at the international Gemini Observatory, a program of NSF's NOIRLab, which is managed by the Association of Universities for Research in Astronomy (AURA) under a cooperative agreement with the National Science Foundation. on behalf of the Gemini Observatory partnership: the National Science Foundation (United States), National Research Council (Canada), Agencia Nacional de Investigación y Desarrollo (Chile), Ministerio de Ciencia, Tecnología e Innovación (Argentina), Ministério da Ciência, Tecnologia, Inovações e Comunicações (Brazil), and Korea Astronomy and Space Science Institute (Republic of Korea).

We thank the referee for his/her extremely helpful and insightful comments that undoubtedly have improved this manuscript.

## ORCID iDs

Mojtaba Taheri  https://orcid.org/0000-0001-7499-5002
Alan W. McConnachie  https://orcid.org/0000-0003-4666-6564
Paolo Turri  https://orcid.org/0000-0002-6451-6239
Davide Massari  https://orcid.org/0000-0001-8892-4301
Giuseppe Bono  https://orcid.org/0000-0002-4896-8841
Giuliana Fiorentino  https://orcid.org/0000-0003-0376-6928
Kim Venn  https://orcid.org/0000-0003-4134-2042
Peter B. Stetson  https://orcid.org/0000-0001-6074-6830